\documentclass{llncs}
\usepackage{xspace}
\usepackage{comment}
\usepackage[
            CJKbookmarks=true,
            bookmarksnumbered=true,
            bookmarksopen=true,
            colorlinks=true,
            citecolor=green,
            linkcolor=blue,
            anchorcolor=red,
            urlcolor=blue
            ]{hyperref}
\usepackage{graphicx,subcaption}
\usepackage{amsmath}

\newcommand{\bpf}{{\em Proof: }}
\newcommand{\epf}{\hfill $\Box$}

\newcommand{\beqa}{\begin{eqnarray}}

\newcommand{\eeqa}{\end{eqnarray}}

\newcommand{\beqan}{\begin{eqnarray*}}

\newcommand{\eeqan}{\end{eqnarray*}}

\newcommand{\beq}{\begin{equation}}

\newcommand{\eeq}{\end{equation}}

\newcommand{\C}{{\cal C}}

\newcommand{\eps}{\varepsilon}

\newcommand{\singlespacing}{\let\CS=\@currsize\renewcommand{\baselinestretch}{0.95}\tiny\CS}

\newcommand{\oneandahalfspacing}{\let\CS=\@currsize\renewcommand{\baselinestretch}{1.25}\tiny\CS}

\newcommand{\doublespacing}{\let\CS=\@currsize\renewcommand{\baselinestretch}{1.39}\tiny\CS}

\newcommand{\be}{\begin{equation}}

\newcommand{\ee}{\end{equation}}

\newcommand{\bc}{\begin{center}}

\newcommand{\ec}{\end{center}}

\newcommand{\bfl}{\begin{flushleft}}

\newcommand{\efl}{\end{flushleft}}

\newcommand{\scheme}{{\sf Prism}\xspace}

\newcommand{\fruitchains}{{\sf Fruitchains}\xspace}

\newcommand{\bitcoinNG}{{\sf BitcoinNG } }

\newcommand{\disccoinnosp}{{\sf DiscCoin}\ignorespaces}

\newcommand{\byzcoinnosp}{{\sf ByzCoin}\ignorespaces}

\newcommand{\thunderella}{{\sf Thunderella }}
\newcommand{\inclusive}{{\sf Inclusive }}
\newcommand{\spectre}{{\sf Spectre }}
\newcommand{\algorand}{{\sf Algorand}\xspace}
\renewcommand{\phantom}{{\sf Phantom}\xspace}
\newcommand{\conflux}{{\sf Conflux }}

\newcommand{\bitcoin}{{\sf Bitcoin}\xspace}
\newcommand{\bitcoinnosp}{{\sf Bitcoin}\ignorespaces}

\newcommand{\ghost}{{\sf GHOST} }

\newcommand{\ps}{{\sf TaiJi}\xspace}
\newcommand{\prism}{{\sf Prism}\xspace}
\newcommand{\streamlet}{{\sf Streamlet}\xspace}

\makeatletter
\renewcommand\subsubsection{\@startsection{subsubsection}{3}{\z@}%
                       {-18\p@ \@plus -4\p@ \@minus -4\p@}%
                       {4\p@ \@plus 2\p@ \@minus 2\p@}%
                       {\normalfont\normalsize\bfseries\boldmath
                        \rightskip=\z@ \@plus 8em\pretolerance=10000 }}

\newcommand{\N}{{\mathcal{N}}}

\renewcommand{\H}{{\mathcal{H}}}
\newcommand{\A}{{\mathcal{A}}}

\newcommand{\V}{{\mathcal{V}}}

\newcommand{\Z}{{\mathcal{Z}}}
\newcommand{\T}{{\mathcal{T}}}
\newcommand{\rmax}{r_{\rm max}}

\newcommand{\kmin}{k_{\rm min}}
\renewcommand{\C}{{\mathcal{C}}}

\newcommand{\E}{{\sf E}}

\newcommand{\cone}{\frac{1-2\beta}{16}}

\makeatother

\renewcommand{\epsilon}{\varepsilon}

\usepackage{algorithm}
\usepackage[noend]{algpseudocode}
\usepackage{xcolor}
\definecolor{azure}{rgb}{0.54, 0.17, 0.89}
\newcommand{\colorcomment}[1]{\Comment{ {\color{azure} #1}} }
\newcommand{\av}[1]{$#1$}
\newcommand{\maincolorcomment}[1]{{\color{azure}// #1 } }

\usepackage{enumitem,kantlipsum}

\usepackage{appendix}

\newcommand{\DAG}{\mathcal{DAG}}

\newcommand{\Hn}{\H_{\textup{notarize}}}
\newcommand{\Hpl}{\H^{\ell}_{\textup{private}}}\newcommand{\Hb}{\H_{\textup{balance}}}

\newcommand{\Hph}{\H^h_{\textup{public}}}
\newcommand{\Hpa}{\H^a_{\textup{public}}}
\newcommand{\Hpds}{\H^{d,\textup{small}}_{\textup{private}}}
\newcommand{\Hpdl}{\H^{d,\textup{large}}_{\textup{private}}}

\newcommand{\dr}{\delta r}
\newcommand{\drtwo}{\delta r'}
\usepackage{amssymb}

\title{\ps: \\Longest Chain Availability with \\
BFT Fast Confirmation}

\author{
Songze Li,
David Tse
\thanks{Email: songzeli8824@gmail.com,
dnctse@gmail.com. }  
}

\institute{ }

\begin{document}
\maketitle

\begin{abstract}
Most state machine replication protocols are either based on the 40-years-old Byzantine Fault Tolerance (BFT) theory or the more recent Nakamoto's longest chain design. Longest chain protocols, designed originally in the Proof-of-Work (PoW) setting, are  available under dynamic participation, but has probabilistic confirmation with long latency dependent on the security parameter. BFT protocols, designed for the permissioned setting, has fast deterministic confirmation, but assume a fixed number of nodes always online.  We present a new construction which combines a longest chain protocol and a  BFT protocol to get the best of both worlds. Using this construction, we design \ps\footnote{\ps:  the Yin of the Longest Chain with the Yang of BFT.}, the first dynamically available PoW protocol which has almost deterministic confirmation  with latency independent of the security parameter. In contrast to previous hybrid approaches which use a single longest chain to sample {\em participants} to run a BFT protocol, our native PoW construction uses many independent longest chains to sample propose {\em actions} and vote {\em actions} for the BFT protocol. This design enables \ps to inherit the full dynamic availability  of \bitcoin, as well as its full unpredictability, making it secure against fully-adaptive adversaries with up to 50\% of online hash power. 

\end{abstract}
\section{Introduction}

\subsection{Background}

Byzantine consensus is a four-decade-old field. Over this long history, many Byzantine Fault Tolerant (BFT) protocols have been developed, with increasing efficiency and simplicity, primarily in the closed permissioned setting with fixed number of authenticated nodes. The invention of \bitcoin by Nakamoto \cite{bitcoin} as the first large-scale consensus protocol in an open {\em permissionless} setting brought several new concepts into the field. First is a new state-machine-replication protocol, the longest chain protocol, an extremely simple protocol with a single action done repeatedly: append a new block to the tip of the longest chain. Second is the use of Proof-of-Work (PoW) \cite{pow} as a lottery to decide who can append the next block.  In addition to sybil-resistance, PoW provides a strong level of unpredictability in the protocol. Third is the notion of {\em dynamic availability}: the protocol continues to run despite unknown and dynamic level of participation from PoW miners; indeed \bitcoin has been running continuously for over a decade during which the total hash power has increased $14$ orders of magnitude. 
Due to these features, Bitcoin is provably secure against a fully adaptive adversary which has less than $50\%$ of the online hash power. (The static hash power case is analyzed in \cite{bitcoin,backbone} in the synchronous round-by-round  model and extended to the $\Delta$-synchronous model in \cite{pss16}; the variable hash power case is considered in \cite{garay_var17,nakamoto_bounded_delay}). The longest chain design proved to be versatile beyond the PoW setting, and  was subsequently adapted to permissioned \cite{sleepy}, Proof-of-Stake (PoS)  \cite{snowwhite,david2018ouroboros,badertscher2018ouroboros}, Proof-of-Space \cite{cohen2019chia} and many other settings, supporting varying degrees of dynamic availability and unpredictability in these settings as well. 

One key feature of longest chain protocols which distinguishes them from BFT protocols is that transaction confirmation is {\em probabilistic}. In longest chain protocols, a transaction  is confirmed if it is in a block $k$-deep in the longest chain. However, there is always a non-zero probability that the block will be removed from the ledger due to the adversary publishing a longer chain in the future. In \bitcoin, probabilistic confirmation is necessitated by the randomness in the Proof-of-Work lottery: there is always a non-zero probability that the adversary gets lucky and wins many blocks over a short period of time. Adaptation of Nakamoto's longest chain protocol to other settings replaces the PoW lottery by lotteries using other types of resources, but the block proposal process is still random and confirmation is still probabilistic. A given security parameter target specifies a particular target deconfirmation probability, and this determines how large $k$ has to be in the $k$-deep confirmation and in turn determines the confirmation latency. Hence, the confirmation latency is {\em security-parameter-dependent}, and gets longer as the desired level of security increases. For example, in \bitcoin, to achieve a deconfirmation error probability of $10^{-3}$ under an adversary having $30\%$ mining power, one has to wait until the transaction is $25$ blocks deep, translating to an average latency of $250$ minutes. This latency increases to $500$ minutes when the target deconfirmation error probability is $10^{-6}$.

In contrast, confirmation in BFT protocols is usually {\em deterministic}: when a transaction is confirmed, one knows for sure that the transaction will not be removed from the ledger (provided that the assumptions for the security of the protocol hold.) The confirmation latency is constant, independent of the security parameter.  However, these protocols are developed for the permissioned setting, and the number of nodes in the network is assumed to be fixed and always participating in the consensus, i.e. these protocols are not dynamically available. This raises an interesting question:

{\em Are there protocols which simultaneously have BFT-like deterministic confirmation with  constant confirmation latency as well as  Bitcoin-level dynamic availability and unpredictability?}
    
\subsection{Main contribution}
In this paper, we answer this question in the affirmative by presenting a construction which combines the longest chain and the  BFT designs to achieve the desired properties. Using this construction, we design a new Proof-of-Work protocol, \ps, which has \bitcoin-level dynamic availability and unpredictability and, with probability  exponentially small in the security parameter $\kappa$, have the following security properties against a fully-adaptive adversary:
\begin{itemize}
    \item \ps is {\em  consistent}, (Theorem \ref{thm:consistent})
    \item \ps is {\em live} as long as the adversary has less than $50\%$ of the online hash power. The \emph{worst case} expected confirmation latency, worst case over all adversarial attacks and all transactions,  is constant, independent of $\kappa$. (Theorem \ref{thm:live})
\end{itemize}


Like \bitcoin, confirmation in \ps is still probabilistic. This is unavoidable, since the PoW lottery is intrinsically random.  However, the confirmation latency is constant, no longer dependents on the deconfirmation probability. In other words, the deconfirmation probability can be made arbitrarily small without sacrificing latency. In this sense, confirmation in \ps is almost deterministic, like in BFT protocols. 

To explain our construction, we first review earlier approaches to improve the latency of \bitcoin, and then build on them.

\subsection{Earlier approaches}

There have been two main approaches to improve the latency of \bitcoin.

\subsubsection{Native-PoW protocols} 

A naive way to improve the latency of \bitcoin is by increasing the mining rate of blocks. However, this comes at the expense of decreased security \cite{ghost}. Thus, the latency of \bitcoin is {\em security-limited}.
 To increase the mining rate while maintaining security, one line of work (\ghost\!\! \cite{ghost},  \inclusive\!\! \cite{inclusive},  \spectre\!\! \cite{spectre}, \phantom\!\!~\cite{phantom}, \conflux\!\! \cite{conflux}) in the literature has used more complex fork choice rules and added reference links to convert the blocktree into a directed acyclic graph (DAG).  This allows blocks to be voted on by blocks that are not necessarily their descendants.  While \ghost remains secure at low mining rates\cite{kiayias2016trees}, there is a balancing attack by the adversary \cite{ghost_attack,better}, which severely limits the security at high mining rates. Thus, like \bitcoin, \ghost is security-limited. The other protocols \inclusive and \conflux that rely on \ghost inherit this drawback. While  \spectre and \phantom improve  latency,  \spectre cannot provide a total order on all transactions (required for smart contracts) and \phantom does not yet have a formal proof of security. 
 
The challenges facing these high-forking protocols arise from the fact that the DAG is {\em unstructured}, due to the excessive random forking when the mining rate is increased. More recent works overcome these security issues by using instead  {\em structured} DAGs \cite{prism,parallel}. \scheme \cite{prism} is based on a decoupling principle: it performs a cryptographic sortition of the mined blocks into proposer, voter and transaction blocks. The voter blocks are further sortitioned into many independent voter chains, each of which votes on the proposer blocks to confirm them. While the individual mining rate in each voter chain is kept low to maintain security, the overall mining rate of the voter blocks is very high to increase the voting rate and hence to decrease the latency. The proposer blocks in turn refer to the transaction blocks, which allows throughput scaling. \cite{parallel} also uses the idea of parallel chains to speed up the confirmation of transactions, but to achieve latency gain, a transaction has to be put in many of the chains, resulting in a tradeoff between throughput and latency. Both \cite{prism} and \cite{parallel} suffer from the limitation that fast total ordering is not guaranteed: transactions that are public double-spends can only be confirmed at a \bitcoin latency, with latency dependent on the security parameter. Thus, these protocols do not offer BFT-like fast confirmation guarantee for all transactions. 

\subsubsection{Hybrid protocols}

Several protocols use BFT consensus in a PoW setting \cite{byzcoin,hybrid,thunderella,abraham2016solida}.  The common idea is to use  PoW longest chain to elect a committee, which will then order transactions, thus reducing the problem into a permissioned consensus problem to achieve improved throughput and latency. In contrast to \bitcoin and the DAG-based protocols, an additional public-key infrastructure is needed to run these hybrid protocols, in order to give identity to the elected committee members. These protocols are not fully dynamically available, as they will stall if some currently elected committee members go offline when executing the BFT protocol in the future. Moreover, these protocols are  insecure against a \emph{fully-adaptive} adversary, which can corrupt the entire elected committee without costing a significant adversarial budget. Such adaptive corruption is impossible in \bitcoin. 

Two examples of such a hybrid approach are \byzcoinnosp~\cite{byzcoin} and its predecessor \disccoinnosp~\cite{disccoin}. They attempt to address the latency shortcoming of \bitcoinNG \cite{bitcoin-ng} but is proven in \cite{hybrid} to be insecure even for {\em static adversaries} when the adversarial fraction is greater than $25\%$. \emph{Hybrid consensus} \cite{hybrid} has an improved (and proven) security up to $33\%$, against a mildly-adaptive adversary.  Its focus is to achieve {\em responsiveness}, i.e. latency depends only on the actual network delay but not on a  delay bound. However, its worst-case latency is still {\em dependent} on the security parameter as in \bitcoin (although it may be possible to improve this worst-case latency by using a better BFT protocol).
A closely-related protocol called \thunderella  \cite{thunderella} achieves very low latency under optimistic conditions, i.e., when the leader is honest and mildly-adaptive adversaries control less that $25\%$ hash power. However even when the adversarial power is very small, a dishonest leader can keep delaying transactions to the \bitcoin latency (since such delaying behavior is detected by a slow PoW blockchain). 

\subsection{\ps: A Construction based on action-sampling }

\begin{figure}[htbp]
   \centering
   \includegraphics[width=\linewidth]{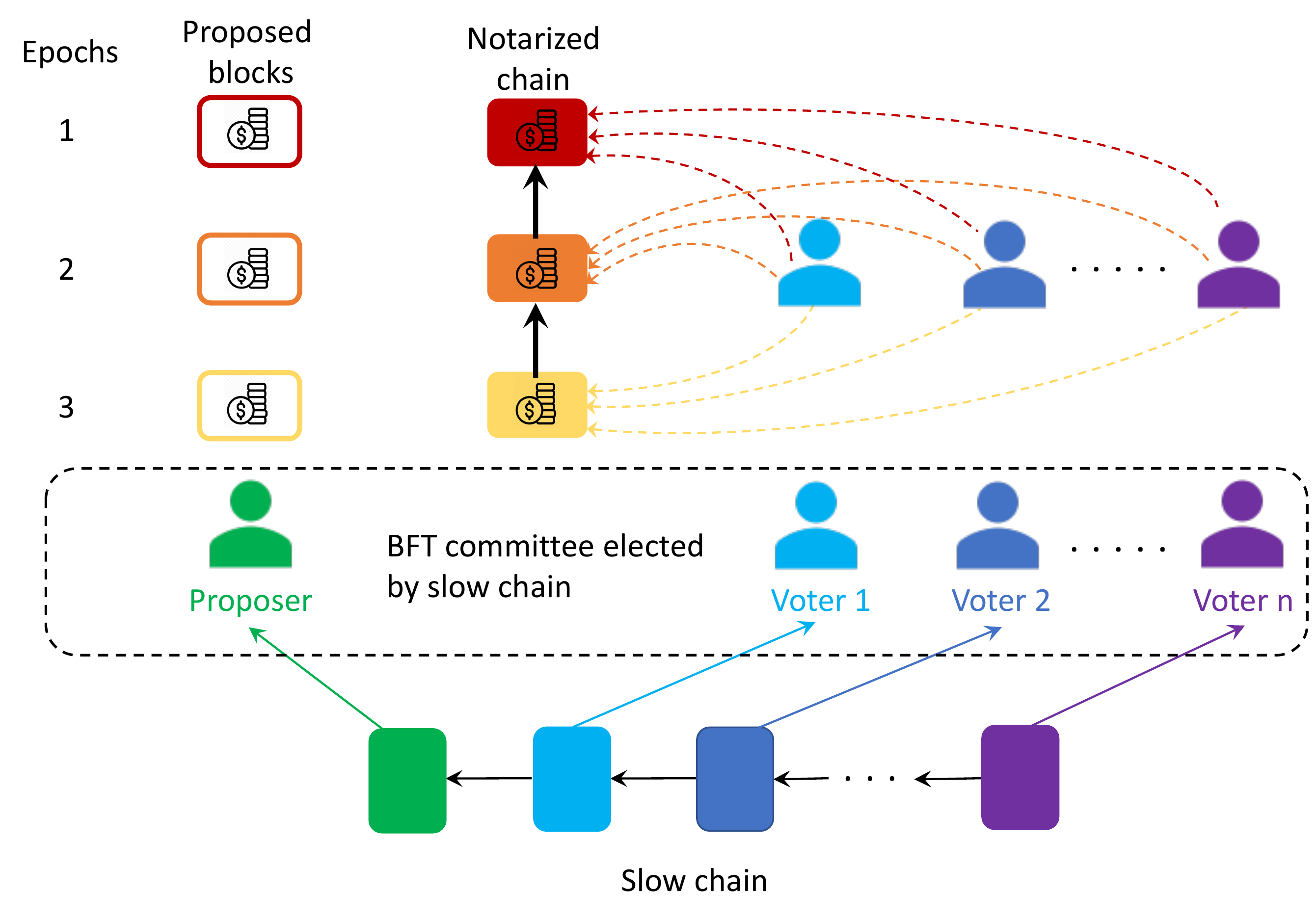} 
   \caption{Hybrid approach: a slow PoW chain elects members of a committee which runs the BFT protocol. Only the miners who win blocks on the main chain are elected. Once the members are elected, the same members will propose and vote over many epochs (say a day), before a new committee is rotated in.}
   \label{fig:phybrid}
 \end{figure}


Our construction is a synthesis of the two approaches above to remove their respective limitations. The result is a native PoW protocol which guarantees fast confirmation latency for {\em all} transactions.


In the hybrid approach, the miners who won blocks on a PoW longest chain (slow chain) are selected to participate in a committee to run a BFT protocol (Fig.~\ref{fig:phybrid}). This two-step approach departs significantly from a basic feature of \bitcoin: miners who won a block has no further special role to play in the protocol compared to any other miner. This departure is evident by the need of a public-key infrastructure (PKI) to give identity to miners to run these hybrid protocols, while \bitcoin has no such need. 

These hybrid protocols use the PoW mechanism in a fundamentally different way than \bitcoin. In \bitcoin, PoW is used as a mechanism to sample {\em actions}, an action being to append a block to the longest chain. In hybrid consensus, PoW is used as a mechanism to sample {\em participants}. This difference leads to loss of the full dynamic availability and unpredictability properties of \bitcoin in these hybrid protocols, because the participants after elected become vulnerabilities: they can be bribed by the adversary or they may go offline stalling the protocol. In contrast, in \bitcoin, a miner is sampled to perform a single action and has no more role to play after the action is performed. Our construction \ps takes the cue from \bitcoin and uses the longest chain to sample actions rather than participants in a BFT protocol, avoiding the need to give identity to miners in running the protocol (hence no PKI required). Yet, we retain the essence of the hybrid approach, which is to exploit the fast latency of BFT protocols.

In \bitcoin, there is only one type of actions: appending a block to the longest chain. This action serves both the role of proposing a block and the role of voting for ancestor blocks in the longest chain. In BFT protocols, the proposing and voting actions are typically separated and so there are two types of actions. \ps uses separate chains to sample proposing and voting actions. This is analogous to \scheme's decoupling of blocks into proposer blocks and voter blocks, each type of blocks in their separate chains. Just like in \scheme, confirmation is sped up in \ps by sampling many votes simultaneously on many independent voter chains; each voter chain can be viewed as a {\em virtual voter}. To ensure that the adversary cannot focus its power to attack the proposer chain or a specific voter chain, cryptographic sortition is used in the mining process so that the type of the mined block is only known after the miner solves the hash puzzle. This technique was first invented in \cite{backbone} and then used in subsequent works, including \fruitchains \cite{fruitchains} and \scheme.  

\begin{figure}[htbp]
   \centering
   \includegraphics[width=\linewidth]{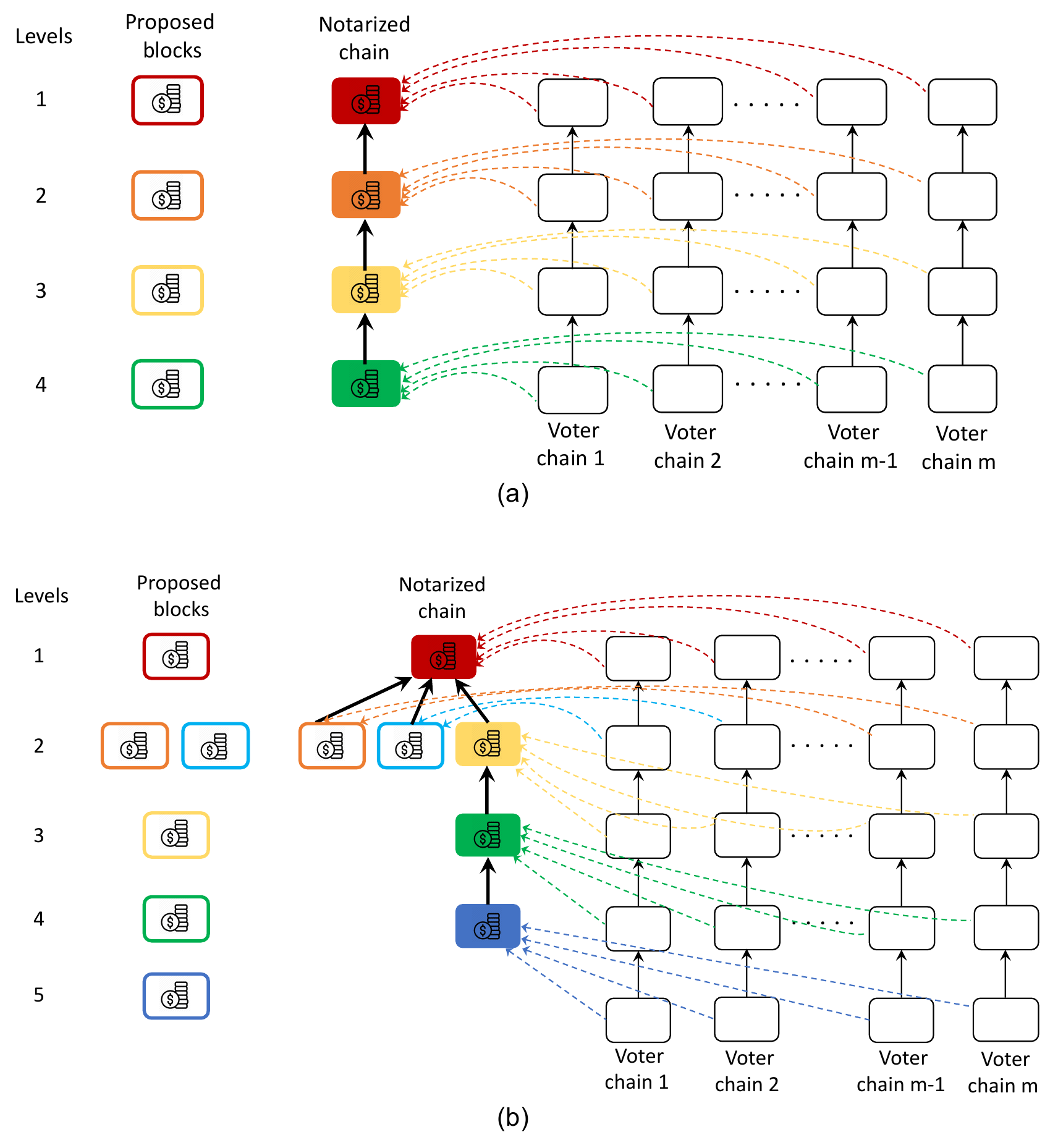} 
   \caption{Schematic of \ps when the BFT protocol is \streamlet. Proposing is done from one proposer tree, level-by-level, and voting is done from $m$ voter chains.  In contrast to Fig. \ref{fig:phybrid}, there is no prior election of proposers and voters but each proposer block or each vote can be won by any miner. (a) When there is no forking, there is one proposer block per level and that same block will be notarized by the voter chains and appear on the notarized chain. (b) The adversary   attacks by releasing a block at the same level as an honest proposer block on level $2$, causing votes to split and resulting in no notarization at that level. However, the unique proposer block at level $3$ will get notarized and the notarized chain continues to grow. This attack is analogous to leader equivocation in classical BFT protocols, and \ps is performing something analogous to leader eviction in this example. In Section \ref{sec:analysis} we show that \ps is in fact secure against all attacks.}  
   \label{fig:parallelize}
 \end{figure}
 
Fig. \ref{fig:parallelize} shows an instantiation of \ps  when the BFT protocol is synchronous \streamlet \cite{streamlet}. In \streamlet, time is divided into epochs of length twice the delay bound. In each epoch, a block is proposed by the randomly chosen epoch leader at the tip of the longest notarized chain. Other nodes then vote on the block, and if the block gets more than $50\%$ of the votes, the block becomes notarized and extends the longest notarized chain. In \ps, the level of the proposer tree plays the role of an  epoch in \streamlet: a miner mines a proposer block on the tip of the proposer tree to propose in the current level.  Simultaneously, it mines on the tip of the longest notarized chain\footnote{Operationally, this means that the hashes of both the parent block in the proposer tree and the parent block in the notarized chain are used in the PoW puzzle for the proposer block.} . Upon solving the hash puzzle, the miner gets to propose a block at the current level and append the same block to the tip of the longest notarized chain. The block is valid regardless of whether it is on the main chain in the proposer tree, as the proposer tree is only used to certify the level of a proposer block.  Like in \streamlet, votes are cast on a proposer block in \ps if the block is on the tip of the longest notarized chain, but in \ps votes are cast from voter chains\footnote{Operationally, a vote is cast on a proposer block from a voter chain if a hash of the proposer block is included in the PoW puzzle for a voter block on the voter chain.} . There is a constraint that each voter chain only votes for one proposer block per level, analogous to an honest node voting only on one proposer block per epoch in \streamlet. A vote is valid if it is on the main chain of its respective voter chain. A proposer block is notarized if it gets sufficient number of valid votes. When three proposer blocks of consecutive levels are notarized consecutively in the notarized chain, then the chain is confirmed until the second of these notarized blocks.

\ps supports three important features:
\begin{enumerate}
    \item {\bf fast latency}: Synchronous \streamlet notarizes a proposer block when it receives $\frac{n}{2} + 1$ votes from the $n$ nodes. However, \ps cannot notarize a block when it receives only $\frac{m}{2}+1$ valid votes from the $m$ voting chains, because some of those votes may be reverted by the adversary forking off the current main chains. However, using results from \cite{prism} it will be shown that, if no other proposer block is made public on the same level, one only needs to wait for a constant amount of time longer to get enough votes to ensure that, with probability exponentially small in the number of voter chains $m$, the block will not get at least $\frac{m}{2} + 1$ votes at all times in the future. By setting $m$ proportional to the security parameter $\kappa$, the target security level is met and at the same time the latency is independent of $\kappa$. This is in contrast to \bitcoin, where the latency has to be proportional to $\kappa$ to average over the randomness of the mining process over time. Here, since we have a large number of voter chains, the averaging is over the many voter chains, whose block mining processes are guaranteed to be independent by the cryptographic sortition process. All we need to guarantee is that a good {\em fraction} of the voter chains are not reverted by the adversary. We don't need to guarantee that {\em every} single voter chain is not reverted. 
    \item {\bf dynamic availability}: Because the proposer blocks and the votes are all sampled from chains, the dynamical availability of \ps follows directly from the dynamic availability of the individual chains. When the total hash power of the network changes, the block generation rate on each of these chains changes, so the effective epoch length changes, but the chains keep on growing and the protocol is live. At a slower time scale, the puzzle difficulty can be adjusted to bring the mining rate back to the target rate, just as in \bitcoin. An important observation is that, due to the cryptographic sortition process, the proposer block generation rate and the vote generation rate is always at a fixed ratio to each other, set by the protocol designer, regardless of the total hash power. This ensures that one can always guarantee that the voting chains cast a  sufficient number of votes per level. 
    \item {\bf unpredictability:} In \ps, the miner who solves the hash puzzle gets to propose a block or vote for a block, but like in \bitcoin it has no further special role after the action is performed. This lack of predictability makes \ps secure against a fully adaptive adversary, just like \bitcoin.
\end{enumerate}

The importance of sampling actions rather than participants to ensure unpredictability has also been  recognized in \algorand \cite{chen2017algorand} in the context of Proof-of-Stake protocols. \algorand samples a small number of voters out of a large population of stakeholders to come to an agreement on a block. To ensure unpredictability, instead of sampling one set of voters for executing all the steps of the Byzantine agreement protocol, \algorand samples an {\em independent} set of voters for each step. A property of {\em player-replaceability} is needed for the Byzantine agreement protocol so that a voter does not need to maintain an internal state from one step to the next. \ps inherits this player-replaceability property from \streamlet. However, there is a crucial difference between the sampling procedures in \algorand and \ps from a dynamic availability point of view. \algorand samples online nodes {\em directly}, and the probability of sampling any particular online node is fixed regardless of how many nodes are online. Hence, the expected number of voters sampled for each protocol step is a fixed {\em fraction} of the number of online nodes. If there are too few online nodes, there will be too few voters and the protocol will stall. This means that \algorand is {\em not} dynamically available. \ps samples online miners {\em indirectly} through sampling the voter chains. Regardless of the total amount of hash power online, each voter chain will eventually vote for a proposer block on each level. Hence, the total number of votes per level is fixed at $m$, the number of voter chains, regardless of the total online hash power. Hence, the protocol never stalls, and \ps is dynamically available. One can think of each of the voter chains as a virtual voter, and the assignment of miners to virtual voters is random and changes over time, depending on which voter blocks are won by which miners. There is a fixed number of such virtual voters, but the rate at which they vote changes with the total online hash power. This rate change is accommodated by changing the rate of proposing proportionally. 


The use of {\em chains} to sample proposing and voting is crucial to enable the dynamic availability of \ps. However, that means the proposing and the voting is no longer organized into time epochs like in \streamlet. This lack of time epochs leads to significant complications in the analysis of the liveness of \ps. In particular, the adversary can release proposer blocks that were  mined privately and earlier to attempt to disrupt liveness. This cannot happen in \streamlet because honest nodes only vote for proposer blocks proposed in the current epoch. Nevertheless we develop new techniques to prove the liveness of \ps as long as the adversary has less than $50\%$ hash power, the same guarantee as \bitcoin.

\subsection{Comparison with Prism} 
\begin{figure}
   \centering
   \includegraphics[width=\linewidth]{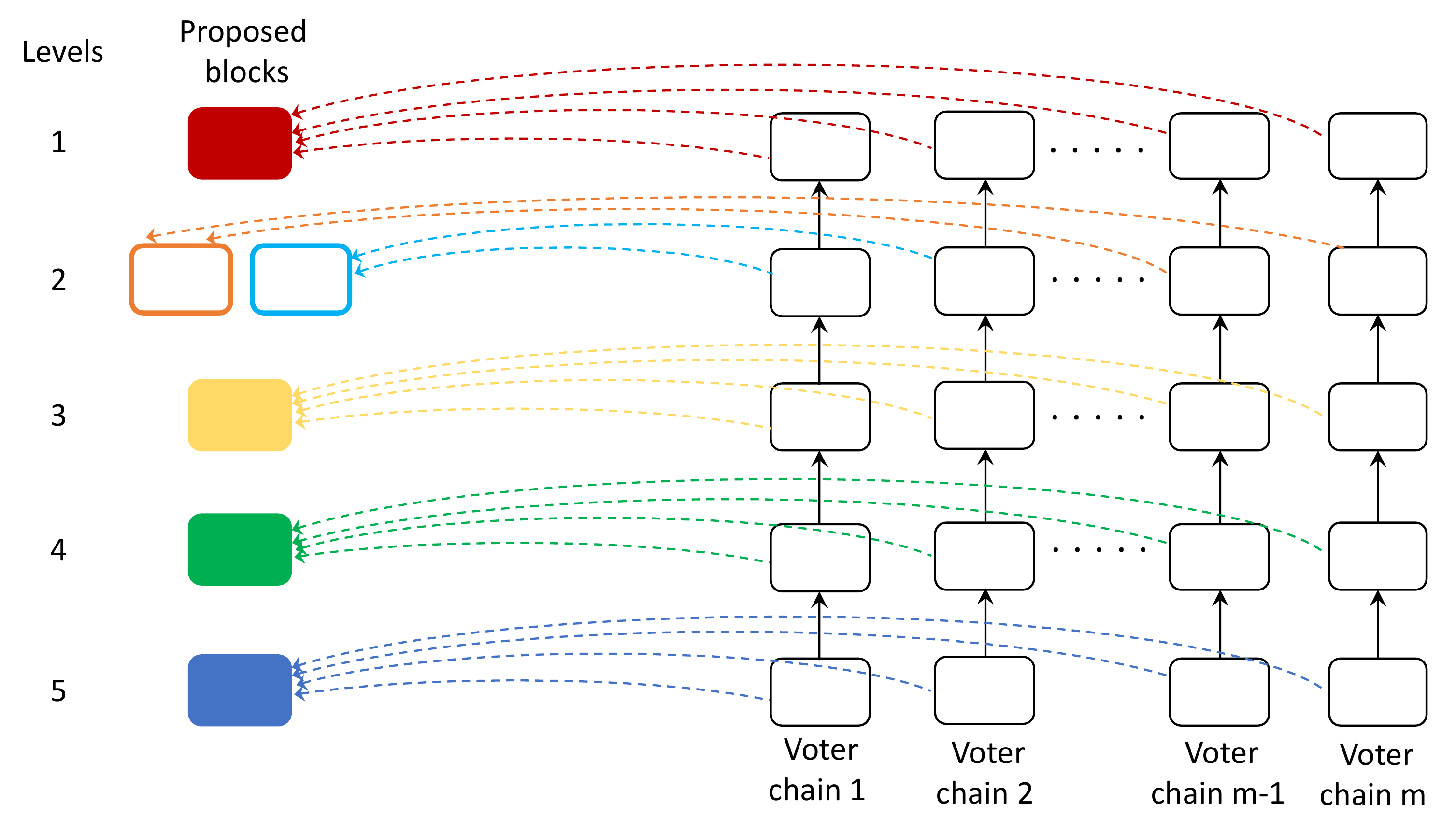} 
   \caption{Execution of \scheme under the same attack scenario as in Figure \ref{fig:parallelize}(b). The orange and blue proposer blocks on level $2$ cannot be fast confirmed by \scheme because the adversary splits the votes they receive.}  
   \label{fig:prism_compare}
 \end{figure}

The use of proposer and voter chains in \ps is inspired by \scheme, but there are important differences between the two protocols. Fig. \ref{fig:prism_compare} gives one such example. Unlike \ps, \prism confirms proposer blocks level-by-level separately, when one of the proposer blocks is guaranteed to have more votes than other blocks,  and does not put proposer blocks in a notarized chain. Hence, the adversary can split the votes evenly between two proposer blocks in the same level and delay their confirmation until every voter chain is irreversible, i.e. \bitcoin latency. As seen in Fig. \ref{fig:parallelize}(b), the adversary cannot do that in \ps because neither proposer block will be notarized and a unique honest proposer block from the next level will be notarized instead, continuing the notarized chain. The two proposer blocks appearing at the same level is equivalent to leader equivocation in BFT protocols, and here we see that \ps, like any good BFT protocol, is able to evict the equivocating leader very quickly. Moreover, by putting the proposer blocks in a (notarized) chain, miners can validate transactions before putting a block onto the chain, just like \bitcoin.  Coupled validation cannot be done in \scheme, where ordering and validation is decoupled, and the ledger needs to be sanitized to remove invalid transactions after it is confirmed. Coupled validation is important for preventing spamming attacks, and is also crucial in enabling light clients who verify the validity of a transaction solely based on the existence of this transaction in a confirmed block.

The present paper focuses on latency, but like \scheme, one can scale the {\em throughput} of \ps  by adding transaction blocks and carrying transactions in the transaction blocks rather than in the proposer blocks. The proposer blocks need only store referring hashes to the transaction blocks, keeping them light for fast communication.  

\subsection{Outline}
Section \ref{sec:model} presents our model, which is similar to the synchronous round-by-round model in \cite{backbone}.  In Section \ref{sec:protocol}, we give a pseudocode description of \ps.  The security analysis of \ps is presented in Section \ref{sec:analysis}.


\section{Model}
\label{sec:model}

This model section is taken from \cite{prism} (with minor modifications), since we will build on their results in our security analysis in Section \ref{sec:analysis}. We include it here for completeness and to introduce the notations.

We consider a synchronous, round-based network model similar to that of Garay \emph{et al.} \cite{backbone}.
We define a blockchain protocol as a pair $(\Pi, g)$, where $\Pi$ is an algorithm that maintains a blockchain data structure $\C$ consisting of a set of \emph{blocks}. 

The function $g({\sf tx},\C)$ encodes a \emph{ledger inclusion rule}; it takes in a transaction ${\sf tx}$ and a blockchain $\C$, and outputs  $g({\sf tx},\C)=1$ if ${\sf tx}$ 
is contained in the ledger defined by blockchain $\C$ 
and $0$ otherwise. 
For example,  in  \bitcoinnosp, $g({\sf tx},\C)=1$ iff ${\sf tx}$ appears in any block on the longest chain.
If there are multiple longest chains, $g$ can resolve ties deterministically, e.g., by taking the chain with the smallest hash value.

The blockchain protocol proceeds in  rounds of  $\Delta$ seconds each.
Letting $\kappa$ denote a security parameter, the \emph{environment} $\Z(1^\kappa)$ captures all aspects external to the protocol itself, such as inputs to  the protocol (i.e., new  transactions) or interaction with outputs.

Let $\N$ denote the set of participating nodes.
The set of \emph{honest} nodes $\H\subset \N$  strictly follow the blockchain protocol $(\Pi,g)$.
\emph{Corrupt} nodes $\N \setminus \H$ are collectively  controlled by an adversarial party $\A$. 
Both honest and corrupt nodes interact with a random function $H:\{0,1\}^*\to \{0,1\}^\kappa$ through an oracle ${\sf H}(x)$, which  outputs $H(x)$.
In each round, each node $n \in \N$ is allowed to query the oracle ${\sf H}(\cdot)$ at most $q$ times. The adversary's corrupt nodes are collectively allowed up to $\beta  q |\N|$ sequential queries to oracle ${\sf H}(\cdot)$, where $\beta$ denotes the fraction of adversarial hash power, i.e., $1-\frac{|\H|}{|\N|}= \beta$.\footnote{$\beta$ for {\bf b}ad. Like \cite{backbone}, we have assumed all nodes have the same hash power, but this model  can easily be generalized to arbitrary hash power distributions.}  Like \cite{backbone}, the environment is not allowed to access the oracle. 
These restrictions model the limited hash rate in the system.

In an execution of the blockchain protocol, the environment $\Z$ first initializes all nodes  as either honest or corrupt;
like \cite{backbone}, once the nodes are initialized, the environment can adaptively  change the set $\H$ between rounds, as long as the adversary's fraction of hash power remains bounded by $\beta$. Thus our model captures a powerful \emph{fully-adaptive} adversary.
Thereafter,  the protocol proceeds in  rounds.
In each round, the environment first delivers inputs to the appropriate nodes (e.g., new transactions), and the adversary delivers any messages to be delivered in the current round.
Here, delivery means that the message appears on the recipient node's input tape.
Nodes incorporate the inputs and any messages (e.g., new blocks) into their local blockchain data structure according to protocol $\Pi$.
The nodes then  access the random oracle ${\sf H(\cdot)}$ as many  times as their hash  power allocation allows. Hence, in each round, users call the oracle ${\sf H(\cdot)}$ with different nonces $s$ in an attempt to find a valid proof of work. 
If an oracle call produces a proof of work, then  the node can deliver a new block to the environment. 
Note that the computational constraints on calling oracle ${\sf H(\cdot)}$ include block validation. 
Since each block only needs to be validated once, validation represents a small fraction of computational demands.

Since each node is allowed a finite number  of calls to ${\sf H(x)}$ in each  round,  the number of blocks mined per round is a Binomial random variable. 
To simplify the analysis, we consider a limit of our model as the number of nodes $|\N| \to \infty$.
As $|\N|$ grows, the proof-of-work threshold adjusts such that the expected number of blocks mined per  round remains constant.
Hence, by the Poisson limit theorem, the total number of blocks mined per round converges to a Poisson random variable. 

All messages broadcast to the environment  are delivered by the adversary. 
The adversary has various capabilities and restrictions.  
(1) Any message broadcast by an honest node in the previous round must be delivered by the adversary at the beginning of the current round to all remaining honest nodes. 
However, during delivery, the adversary can present these messages to each honest node in whatever order it chooses. 
(2) The adversary cannot forge or alter any message sent by an honest node.
(3) The adversary can control the actions of corrupt nodes.
For example, the adversary can choose how corrupt nodes allocate their hash power, decide block content, and release mined blocks. 
Notably,  although honest blocks publish mined blocks immediately,  the adversary may choose to keep blocks they mined private and release in some future round.
(4) The adversary can deliver corrupt nodes' messages to some honest nodes in one round, and the remaining honest nodes in the next round. We consider a ``rushing'' adversary that observes the honest nodes' actions before taking its  own action for a given round.
Notice that we do not model rational users who are not necessarily adversarial but nevertheless may have incentives to deviate from protocol. 

\paragraph{Metrics.}
We let random variable ${\sf VIEW}_{\Pi,\A,\Z}$ denote the joint view of all parties over all rounds; here we have suppressed the dependency on security parameter $\kappa$.
The randomness is defined over the choice of function $H(\cdot)$,  as well as any randomness in the adversary $\A$ or environment $\Z$. 
Our goal is to reason about the joint  view for all possible adversaries $\A$ and environments $\Z$. 
In particular, we want to study the evolution of $\C_i^r$, or the blockchain of each honest node $i\in \H$ during round $r$.
Following the \bitcoin backbone protocol model \cite{backbone}, we consider protocols that execute for a finite execution  horizon $\rmax$, polynomial in $\kappa$.  
Our primary concern will be the efficiency of \emph{confirming} transactions.

\begin{definition}
\label{def:epsilon}
We say a transaction ${\sf  tx}$ is  $(\epsilon,\A,\Z,r_0,\kappa)$-cleared iff under an adversary $\A$,  environment  $\Z$, and security parameter $\kappa$, 
$$
 \mathbb P_{{\sf VIEW}_{\Pi,\A,\Z}} \left( \bigcap_{\substack{r \in \{r_0, \ldots, \rmax\}\\  i \in \H }}  \left \{g({\sf tx},\C_i^r)=b\right \}  \right)\geq 1-\epsilon - {\sf negl}(\kappa),
$$
where $b\in\{0,1\}$; $b=1$ corresponds to confirming the transactions and $b=0$ corresponds to rejecting the transaction.
\end{definition}
That is, a transaction is considered confirmed (resp. rejected) if all honest party will include (resp. exclude) it from the ledger with probability more than $\epsilon$ plus a term negligible in $\kappa$ resulting from hash collisions, which we ignore in our analysis. 
We suppress the notation  $\kappa$ from here on.

Our objective is to optimize the latency of confirming transactions in a blockchain protocol.
We let $\T$  denote the set of all transactions generated during the execution horizon. 

\begin{definition}[Latency]
For a transaction ${\sf  tx}$, let $r({\sf tx})$ denote the round in which the transaction was first introduced by the envioronment, and let random variable $ R_{\epsilon}({\sf tx})$ denote the smallest round $r$ for which ${\sf  tx}$ is $(\epsilon,\A,\Z,r)$-cleared.
The worst-case expected $\epsilon$-\emph{latency} 
is defined as:
\begin{equation}
    \begin{aligned}
 \tau_\epsilon &\triangleq       &\underset{ \Z, \A, {\sf tx} \in \T}{\max}& & \mathbb{E}_{{\sf VIEW}_{\Pi,\A,\Z}} \left [ R_\epsilon({\sf tx}) -  r({\sf tx}) \right]
    \end{aligned}
    \label{def:latency}
\end{equation}
\end{definition}

Note that if $\tau_\epsilon$ is finite, 
it implies that the blockchain has both consistency and liveness properties. 

\section{Protocol description}
\label{sec:protocol}

\begin{figure}[htbp]
\includegraphics[width=\linewidth]{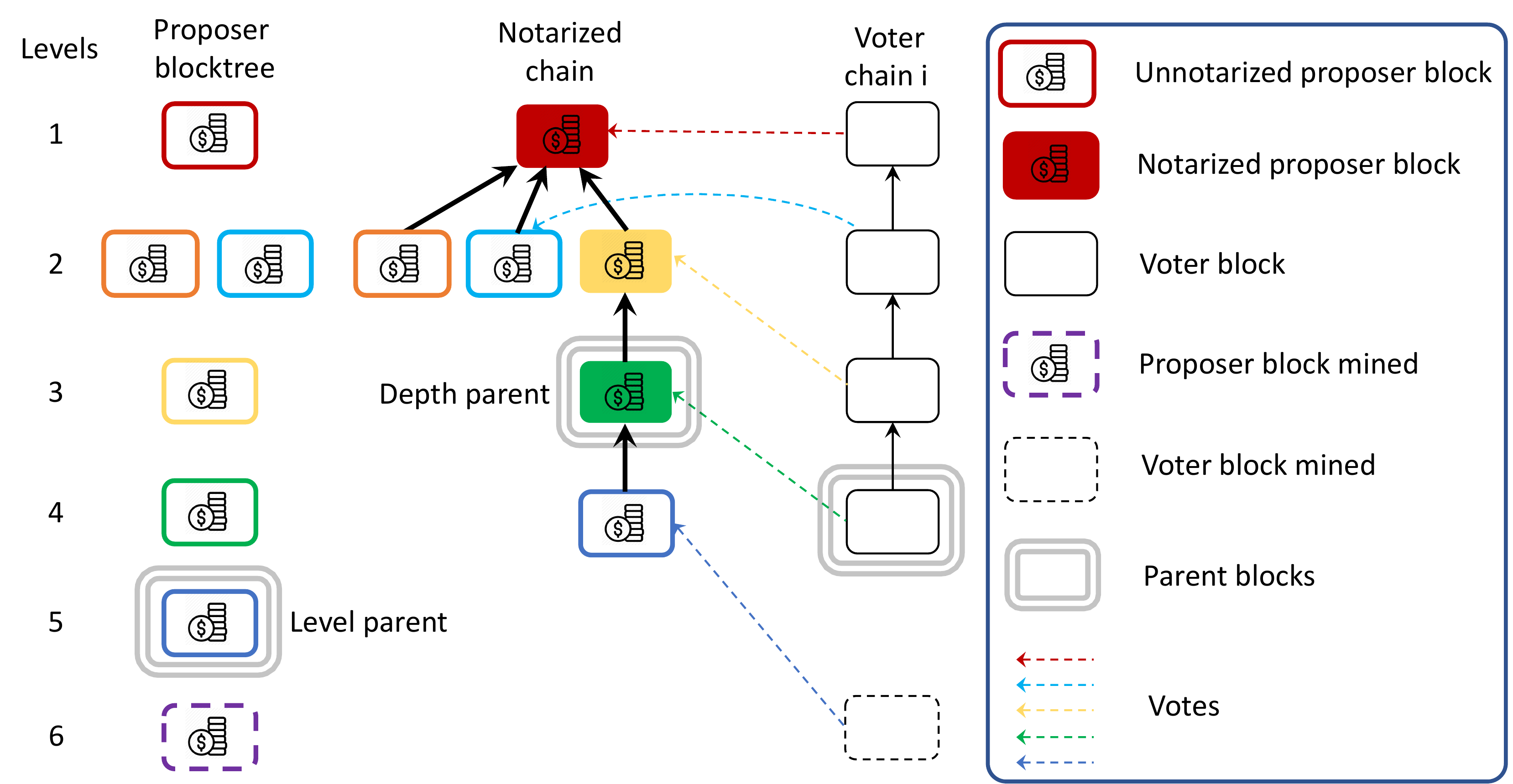}
   \caption{Snapshot of a miner's proposer blocktree and notarized chain: the previously mined blocks have solid boundary whereas blocks which are being mined have dotted-boundary. Notarized proposer blocks are made solid and become a part of the notarized chain. A miner simultaneously mines on a level parent on proposer blocktree, a depth parent on notarized chain, and a parent on voter block blocktree $i$, $\forall i =1,2,\ldots,m$.}
   \label{fig:prism_mining}
 \end{figure}

We first describe the content and roles of the two types of blocks in the $\ps(\Pi, g)$ blockchain, proposer blocks and voter blocks.  
We then present Algorithm \ref{alg:prism_minining}, which defines the protocol $\Pi$ and the blockchain data structure $C$. 
We then define the \textit{ledger confirmation rule}, $g$, in Algorithm \ref{alg:prism_con}. 
\ps's blockchain data structure, $C$, contains one proposer blocktree, one notarized proposer blockchain, and $m$ voter blocktrees, as shown in Fig. \ref{fig:parallelize}. 
We have two distinct types of blocks:

\textit{Proposer blocks:} proposer blocks contain transactions that are proposed to be included in the ledger, and constitutes the skeleton of the \ps blockchain. Proposer blocks are simultaneously mined on the proposer blocktree and the notarized blockchain, both following the longest-chain rule. The \textit{level} of a proposer block is defined as its distance from the proposer genesis block on the proposer blocktree. The level of a proposer block is analogous to the concept of ``epoch'' in BFT protocols, which roughly indicates the time when a block was first proposed. The \emph{depth} of a proposer block, on the other hand, is its distance from the proposer genesis block on the notarized proposer blockchain. Among all proposer blocks on the proposer tree, only notarized ones can be added to the notarized chain. For instance, in Fig. \ref{fig:prism_mining}, yellow proposer block was mined on level $3$ of the proposer blocktree and depth $2$ of the notarized chain. It becomes a part of the notarized chain after notarization (i.e., turning solid). On the notarized chain, our protocol selects a prefix for confirmation, and the transactions in the confirmed blocks are used to construct the output ledger. 

 \begin{algorithm}[H]
{\fontsize{8pt}{8pt}\selectfont \caption{\ps: Mining}\label{alg:prism_minining}
\begin{algorithmic}[1]

\Procedure{Main}{ }
    \State \textsc{Initialize}()
    \While{True}
        \State $header, Ppf, Cpf$ =  \textsc{PowMining}() \label{code:blockBody1}
        \State \maincolorcomment{Block contains header, parent, content and Merkle proofs}
        \If{header is a \textit{prop block}}
        \State $block \gets \langle header, lvlParent, depParent, txPool, Ppf, Cpf\rangle$\label{code:blockBody2}
        \ElsIf{header is a \textit{block in voter} blocktree $i$}
        \State $block \gets \langle header, vtParent[i], votesOnPrpBks[i], Ppf,\label{code:blockBody3} Cpf\rangle$
        \EndIf
        \State \textsc{BroadcastMessage}($block$)  \colorcomment{Broadcast to peers}

    \EndWhile
\EndProcedure

\vspace{1mm}
\Procedure{Initialize}{ } \colorcomment{ All variables are global}
\vspace{0.5mm} \State \maincolorcomment{Blockchain data structure $C  = (prpTree, notChain, vtTree) $}
\State $prpTree \gets genesisP$ \label{code:prpGenesis} \colorcomment{Proposer Blocktree}
\State $notChain \gets genesisP$ \label{code:notGenesis} \colorcomment{Notarized Blockchain}
\For{$i \gets 1 \; to \; m$}
\State $vtTree[i] \gets genesisM\_i$ \colorcomment{Voter $i$ blocktree} \label{code:voterGenesis}
\EndFor
\State \maincolorcomment{Parent blocks to mine on} \label{code:VarprpParent}
\State $lvlParent$ $\gets genesisP $ \colorcomment{ Proposer block to mine on in proposer blocktree}
\State $depParent$ $\gets genesisP $ \colorcomment{ Proposer block to mine on in notarized chain}
\For{$i \gets 1 \; to \; m$}
\State $vtParent[i]$ $\gets genesisM\_i $ \colorcomment{Voter tree $i$ block to mine on}
\EndFor \label{code:VarvtParent}
\vspace{00.4mm}\State \maincolorcomment{Block content} \label{code:allcontent}
\State  $txPool$ $\gets \phi$ \colorcomment{Prp block content: Txs to add in prp blks} \label{code:txblockcontent}
\For{$i \gets 1 \; to \; m$}
\State $votesOnPrpBks(i) \gets \phi$ \colorcomment{Voter tree $i$ blk content } \label{code:vtblockcontent}
\EndFor
\EndProcedure

\vspace{1mm}
\Procedure{PowMining}{ }
\While{True}
\State \maincolorcomment{Assign content for all block types/trees}
\For{$i \gets 1 \;  to \; m$} $vtContent[i] \gets votesOnPrpBks[i]$ \label{code:voteIneffcient}
\EndFor
\State $prContent \gets txPool$
\State \maincolorcomment{Define parents and content Merkle trees}
\State $parentMT\gets$MerkleTree($vtParent,lvlParent,depParent$) 
\State $contentMT\gets$MerkleTree($vtContent,prContent$) \label{code:endContent}
\State nonce $ \gets $ RandomString($1^\kappa$)
\State \maincolorcomment{Header is similar to Bitcoin}
\State header $\gets \langle$ $parentMT.$root, $contentMT.$root, nonce $\rangle$
\vspace{00.4mm} \State \maincolorcomment{Sortition into different block types/trees } 
\If {Hash(header)  $\leq mf_v$}\label{code:sortitionStart} \colorcomment{Voter block mined}
    \State $i\gets  \lfloor $Hash(header)/$f_v\rfloor$
    \textbf{and} \textit{break} \colorcomment{on tree $i$ }
\ElsIf{ $mf_v< $ Hash(header) $\leq mf_v+f_p$} 
    \State  $i \gets m+1$ \textbf{ and} \textit{break}\colorcomment{Prop block mined}
\EndIf
\EndWhile \label{code:sortitionEnd}
\maincolorcomment{Return header along with Merkle proofs}
\State \Return $\langle header, parentMT.$proof($i$),  $contentMT.$proof($i) \rangle$ \label{code:minedBlocm}
\EndProcedure

\vspace{1mm}
\Procedure{ReceiveBlock}{\textsf{B}} \colorcomment{Get block from peers}    

\If{\textsf{B} is a valid \textit{block on $i^{\text{th}}$ voter tree}} \label{code:vtblk_r1}
    \State $vtTree[i]$.append(\textsf{B}) \textbf{and} $vtTree[i]$.append(\textsf{B}.ancestors())    
    \If{\textsf{B}.chainlen $> vtParent[i]$.chainlen} 
    \State $vtParent[i] \gets \textsf{B} $ \textbf{and} $votesOnPrpBks$($i$).update(\textsf{B}) \label{code:updateVoteMine}
    \EndIf\label{code:vtblk_r2}

\ElsIf{\textsf{B} is a valid \textit{prop block}} \label{code:prpblk_r1}
    \If{\textsf{B}.level $==prpParent$.level+$1$}
    \State $prpParent \gets \textsf{B}$ \label{code:updatePropToMine}
    \For{$i \gets 1 \;  to \; m$}  
    \colorcomment{Add vote on level $\ell$ on all $m$ trees}
    \State $votesOnPrpBks(i)[\textsf{B}.level] \gets \textsf{B}$   \label{code:addVote} 
    \EndFor
    \ElsIf{\textsf{B}.level $ > prpParent$.level+$1$}
        \State \maincolorcomment{Miner doesnt have block at level $prpParent$.level+$1$}
        \State \textsc{RequestNetwork(\textsf{B}.parent)} 
    \EndIf
    \State  $prpTree[\textsf{B} $.level].append(\textsf{B})

\EndIf
\EndProcedure

\vspace{1mm}
\Procedure{ReceiveTx}{\textsf{tx}} 
\If {\textsf{tx} has valid signature }   $txPool$.append(\textsf{B})
\EndIf    
\EndProcedure
\end{algorithmic}
}
\end{algorithm}

    \textit{Voter blocks:}
    Voter blocks are mined on $m$ separate voter blocktrees, each with its own genesis block, according to the longest-chain rule. 
    We say a voter block \emph{votes} on a proposer block $B$ if it includes a pointer to $B$ in its payload.
    Note that unlike many BFT protocols, a malicious miner in \ps cannot equivocate when voting because voter blocks are sealed by proof of work.
    Even if a miner mines conflicting voter blocks and tries to send them to disjoint sets of honest users, all users will receive both blocks within one round.
    Each longest chain from each voter blocktree can cast at most one vote for each level in the proposer blocktree.
    More precisely, a voter block votes on blocks that simultaneously satisfy 1) on levels in the proposer tree that are greater than the levels voted by the voter block's ancestors, and 2) extending the longest notarized chain currently seen by the miner of the voter block.
    Fig. \ref{fig:prism_mining} shows voter blocktree $i$ and its votes (dotted arrows) on each level of the proposer blocktree.
    A proposer block that received a certain number of votes is considered notarized. The number is set that on each level of the proposer tree at most one block can be notarized.

The process by which a  transaction is included in the ledger is as follows: 
1) the transaction is included in a proposer block $B$;
2) $B$ is notarized;
3) $B$ is confirmed, either itself or one of its notarized descendants meets the confirmation criteria.

\subsection{ Protocol $\Pi$}
Algorithm \ref{alg:prism_minining} presents \ps's protocol $\Pi$. 
The protocol begins with a trusted setup, in which the environment generates  genesis blocks for the proposer blocktree (also servers as the Genesis block of the notarized blockchain) and each of the $m$ voter blocktrees.
Once the trusted setup completes, the protocol enters the mining loop.  

Whereas \bitcoin miners mine on a single blocktree, \ps miners simultaneously mine one proposer block and $m$ voter blocks. Each proposer block has two parent proposer blocks, one as the ``level'' parent who is on the greatest level of the current proposer tree, and another one as the ``depth'' parent who is the tip of the longest notarized proposer chain. In Fig.~\ref{fig:prism_mining}, the upcoming purple proposer block has the blue block as its level parent, and the notarized green block as its depth parent. Each voter block on voter tree $i$ has the tip of the longest chain in voter tree $i$ as its parent. This simultaneous mining happens via cryptographic sortition.
Roughly, a miner first generates a  ``superblock'' that contains enough information for all $m+1$ blocks simultaneously.
It then tries different nonce values; upon mining a block, the output of the hash is deterministically mapped to either a voter block (in one of the $m$ trees), or a proposer block. As shown in lines 38-41 in Algorithm \ref{alg:prism_minining}, protocol designer can adjust the lengths of the hash value intervals (e.g., $f_p$ for proposer block and $f_v$ for a voter chain) to control the rate of generating proposer and voter blocks.
\begin{definition}\label{def:mine_rate}
Here we define, as design parameters, the average rate of mining proposer blocks as $\bar{f}_p$ blocks/round, and the average rate of mining voter blocks on each voter chain as $\bar{f}_v$ blocks/round.
\end{definition}
After sortition,  the miner discards unnecessary information and publishes the block  to the environment.

While mining, each miner maintains outstanding content for each of the  $m+1$ possible mined blocks.  
In \bitcoinnosp, this content would be the transaction memory pool, but since \ps has two types of blocks, each miner stores different content for each block type.
For proposer blocks, the content consists of all transactions that have not been confirmed.
For voter blocks, the content are pointers to a list of proposer blocks selected according to the following voting rule.
\begin{definition}[Voting rule]
For a voter block on the $i$th voter tree, it first identifies a set of proposer blocks that simultaneously satisfy
\begin{enumerate}
     \item With depth one greater than that of the current tip of the longest notarized chain;
    \item At some level of the proposer blocktree that is greater than all levels voted so far by the ancestors of the voter block on the longest chain in the $i$th voter tree.
\end{enumerate}
Within such identified set, for each distinct level of the proposer blocks, the voter block includes a pointer (vote) to the proposer block that was received earliest at the miner.
\end{definition}

To illustrate this voting rule, as shown in Fig. \ref{fig:prism_mining}, the bottom voter block being mined votes for the blue proposer block as it is one level greater than the green block voted previously on voter chain $i$, and it is extending the tip of the notarized chain (green block). 
Upon collecting this content for potential proposer or voter blocks, the miner starts to mine a block.  
Instead of naively including all the $m+2$ parents and content hashes in the block, \ps's header contains a) the Merkle root of a tree with $m+2$ parent blocks, b) the Merkle root of a tree with $m+1$ contents, and c) a nonce.
Once a valid nonce is found, the block is sortitioned into a proposer block or a voter block on one of the $m$ voter trees. 
The mined, sortitioned block consists of the header, the appropriate parent(s) and content, and their respective Merkle proofs.
For instance, if the mined block is a proposer block, it would contain two proposer parent references, proposer content (transactions), and appropriate Merkle proofs.

While mining, nodes may receive blocks from  the network, which are processed in  much the same way as \bitcoinnosp.
Upon receiving a new block, the  miner first checks validity. For a block $B$ to be considered valid, it has to satisfy the PoW inequality and the miner has all the blocks locally (directly or indirectly) referred by $B$.  
If the miner lacks some referred blocks, it requests them from the network. For a proposer block, the miner also needs to check the referred depth parent was indeed notarized in its local view. For a voter block $V$, the miner also needs to check that the proposer block it voted for has 1) a level that is greater than all proposer blocks voted by the ancestors of $V$, and 2) a depth that is no less than all proposer blocks voted by ancestors of the $V$.
Upon receiving a valid voter block, the miner updates the longest voter chain if needed, and updates the vote counts accordingly.
Upon  receiving a valid proposer block $B$ at a level $\ell$ greater than the previous greatest level, the miner makes $B$ the  new  level parent for future proposer blocks.

\subsection{Ledger confirmation rule $g$}
\label{sec:confirmation}

\begin{algorithm}[H]
{\fontsize{8pt}{8pt}\selectfont \caption{\ps: Tx confirmation}\label{alg:prism_con}
\begin{algorithmic}[1]

\Procedure{IsTxConfirmed}{$tx$, $notaBks$}\label{code:fastConf}
\colorcomment{$notaBks$ is the set of notarized blocks}
\State $ledger$ = \textsc{BuildLedger}($notaBks$)
\State \Return $tx$ is in \av{ledger} \colorcomment{Return True if tx is included in the output ledger} \label{code:fastConfirmTx}
\EndProcedure

\vspace{1mm}
\Procedure{IsBlkNotarized}{$prpBk$}\label{code:notarize}
        \State $votesNdepth \gets \phi$ 
        \For{$i$ in $1\;to\;m$} \label{code:getvotes_start}
        \State  $votesNdepth[i] \gets \textsc{GetVoteNDepth}(i, prpBk)$ 
        \EndFor
        \Return {IsPropBlkNotarized($votesNdepth$)}\colorcomment{Refer Definition \ref{def:notarization}}
\EndProcedure

\vspace{1mm}
\State \maincolorcomment{Return the vote of voter blocktree $i$ for $prpBk$ and  depth of the vote}
\Procedure{GetVoteNDepth}{$i, prpBk$}
    \State $voterMC \gets vtTree[i].LongestChain()$
    \For{$voterBk$ in $voterMC$} \label{code:voteCountingStart}
        \For{$bk$ in $voterBk$.votes}
            \If{$bk$$ == prpBk$} 
                    \State \maincolorcomment{Depth is \#of children bks of voter bk on main chain}
\State  \Return $voterBk$.\text{depth} \label{code:voteCountingEnd}
            \EndIf
        \EndFor
    \EndFor
\EndProcedure
\vspace{1mm}
\Procedure{BuildLedger}{\av{notaBks}} \colorcomment{Input: list of notarized proposer blocks}\label{code:buildLedgerFunc}
\State \av{ledger} $\gets []$ \colorcomment{List of valid transactions}
\State \av{confChain} $\gets []$ \colorcomment{Confirmed proposer chain}
\For{\av{bk} in \av{notaBks}} 
\If{\av{bk} is confirmed}  \colorcomment{Refer Definition~\ref{def:final}}
\State \av{bkChain} $\gets$ Confirmed chain ending in bk
\If{\av{bkChain.length()} $>$ \av{confChain.length()}}
\State \av{confChain} $=$ \av{bkChain}
\EndIf
\EndIf
\EndFor
\For{\av{bk} in \av{confChain}}
\State \av{ledger}.append(\av{bk}.\av{txs})
\EndFor
\State \Return \av{ledger}
\EndProcedure

\end{algorithmic}
}
\end{algorithm}

\noindent {\bf Notarization.} A proposer block in level $\ell$ is said to be \emph{notarized} if it is expected to receive more votes than other proposer blocks on the same level. More precisely, as stated in the subroutine \textsc{IsBlkNotarized()} in Algorithm \ref{alg:prism_con},  
a proposer block is notarized when its current votes, discounted by possible future loss of votes due to change of longest chains on some voter trees, exceeds $\frac{m}{2}$. The precise definition of the notarization rule is given in Definition \ref{def:notarization}.

We say a chain of proposer blocks, connected through references to depth proposer parents starting from the proposer Genesis block, is a \emph{notarized chain} if all the blocks on the chain are notarized. 

Once a miner observes the notarization of some proposer block $B$ at a depth that is larger than the depths of all other notarized blocks, the miner marks $B$ as the tip of the notarized chain and the new depth parent for future proposer blocks.  

\begin{definition}[Confirmation]\label{def:final}
When a node sees three adjacent blocks on a notarized proposer chain that are also on three consecutive levels of the proposer tree, it confirms the second of the three blocks, together with its notarized prefix chain.
\end{definition}

The formation of the output ledger is rather straightforward. All transactions in the proposer blocks on the confirmed chain are confirmed and will be included in the output ledger. The ordering of the confirmed transactions are derived from the chaining of the confirmed proposer blocks, and the ordering of the transactions within each proposer block.

\section{Security Analysis}
\label{sec:analysis}

\subsection{Design parameters of \ps}
The design parameters of \ps are the number of voter chains $m$, the proposer block mining rate $\bar{f}_p$ per round and the voter block mining rate $\bar{f}_v$ per round per chain.  We will choose these parameters such that \ps is secure against an adversary with $\beta$ fraction of hash power with probability of loss of security exponentially small in the security parameter $\kappa$, over a finite horizon of $\rmax$ rounds. $\rmax$ is a polynomial function of $\kappa$. In particular, we will choose the voter block mining rate $\bar{f}_v$ to be small so that there is little forking in each of the voter chains. This will ensure a security level $\beta$ of close to $1/2$ for the protocol. 

From these parameters, we will define:
\begin{eqnarray}
    \gamma &\triangleq& \frac{1}{36}(1-2\beta)^2, \;\; c_1 \triangleq \cone,  \nonumber\\ 
    \kmin &\triangleq& \frac{4}{\gamma }\log \frac{200}{\gamma c_1}, \nonumber\\
     \delta_k & \triangleq&  \max\left(\frac{c_1}{1+2k}, \frac{(1-2\beta)c_1}{1+32\log m} \right), \nonumber\\
    \eps_m &\triangleq & r_{\text{max}}^2e^{-\frac{(1-2\beta)c_1m}{2+64\log m}}, \label{eqn:constants}
\end{eqnarray}
which will be used in the analysis.

\subsection{Basic properties of notarized blocks}
We first define in the following three critical attributes of a proposer block in \ps.
\begin{definition}
We define $\ell(B)$, $d(B)$, and $r(B)$, for a proposer block $B$, as the level of $B$ in the blocktree, the depth of $B$ on the notarized blockchain, and the round in which $B$ is mined. 
\end{definition}

Notarization of proposer blocks in \ps comes from the voter chains. The behavior of these chains are well-studied in \cite{prism}, and we will use the results from there extensively. In particular, the key properties studied in \cite{prism} are macroscopic versions of the basic chain-growth (Lemma E.2) , common-prefix (Lemma E.3) and chain-quality (Lemma E.4) properties, taken across the entire ensemble of voter chains. Chain-growth, common-prefix and chain-quality were properties introduced in \cite{backbone} for a single longest chain. Lemmas E.2, E.3 and E.4 in \cite{prism} show that under a typical event $\texttt{T}$ on the mining times of the voter blocks, a large fraction of the voter chains satisfies these three properties. The typical event $\texttt{T}$ is defined in Lemma E.1, where it is shown that the probability of $\texttt{T}$ is at least $1-\eps_m$, with $\eps_m$ defined above and goes to zero with the number of voter chains $m$ exponentially. All the results in this section is conditional on the typical event $\texttt{T}$. Importantly, this event $\texttt{T}$ only depends on the voter blocks mining process, and is independent of the proposer block mining process. The event $\texttt{T}$ can be viewed as a macroscopic version of the notion of typical execution introduced in \cite{backbone} for a single chain.

Let $V_B[r]$ be the number of votes for proposer block $B$ in round $r$.  A vote is at least {\em $k$-deep} if it is from a voter block that is on a longest chain and has $k-1$ or more descendants on the longest chain.  Let $V^k_B[r]$ be the number of votes which are at least $k$-deep for proposer block $B$ at round $r$. Define:
\begin{align}
 \underbar{V}_B[r] \triangleq \max_{k \geq k_{\rm min}} \left (V_B^{k}[r] -\delta_k m \right )_{+},
\end{align}

\begin{lemma} (Lemma E.5 in \cite{prism})
\label{lem:votes_bounds}
Under the typical event $\texttt{T}$, the number of votes on any proposer block $B$ in any future round $r_f\geq r$, $V_B[r_f]$, satisfies
\begin{equation}
 V_B[r_f] \ge    \underbar{V}_B[r] ,
 \nonumber
\end{equation}
\end{lemma}

Lemma \ref{lem:votes_bounds} says that $\underbar{V}_B[r]$ is a high-probability lower bound on how many of the current votes will stay in the longest voter chains forever. Some of the current votes may be reversed by the adversary, but the deeper they are in the voter chains, the less likely they will be reversed. The proof of Lemma \ref{lem:votes_bounds} is based on a {\em macroscopic common prefix property}, which bounds the {\em fraction} of voter chains that will violate the common prefix property as a function of the depths of the votes. Details of the proof can be found in \cite{prism}.

Since $\underbar{V}_B(r)$ is a lower bound on the future votes, and it can be computed from the current blockchain, it can be used as the notarization criterion.

\begin{definition}\label{def:notarization}
A proposer block $B$ is said to be {\em notarized} at round $r$ if 
\begin{equation}
\underbar{V}_B[r] \ge \frac{m}{2} + 1.    
\end{equation}
\end{definition}




\begin{lemma}
\label{ass:half_votes}
Under typical event $\texttt{T}$, it holds that when any proposer block  is notarized, it will get at least $\frac{m}{2} + 1$ votes in all future rounds.
\end{lemma}

\bpf
Follows immediately from Lemma \ref{lem:votes_bounds}.
\epf

Next, we will show that a proposer block becomes notarized after a finite number of rounds if there are no competing blocks. This is crucial for proving the liveness of the protocol. 





\begin{lemma}
\label{lem:notarize}
Assume  $\beta< 1/2$, $\bar{f}_v$ is chosen sufficiently small, and $m$ sufficiently large. Under the typical event $\texttt{T}$, the following statement is true for all honest proposer blocks: there exists a $\dr$, depending only on $\beta$,  such that if an honest proposer block is broadcast at round $r$ and if by round $r+\dr$, there is no other block made public at the same 
level or notarized to advance the length of the longest notarized chain at round $r$ by one, then:
\begin{enumerate}
    \item The honest proposer block is notarized  in all future rounds;
    \item More than $m/2$ of the voting chains vote for the honest block in all future rounds.  
\end{enumerate}
\end{lemma}

\bpf
Let $H$ be a honest proposer block that appears at round $r$. It is on the tip of the longest notarized chain. It follows from Lemma E.2 (macroscopic chain-growth), Lemma E.4 (macroscopic chain-quality) in \cite{prism} and $\beta < 1/2$ that a large fraction of voter chains grow and have positive chain quality. Hence, if one waits for a sufficient, but finite, many number of rounds, one can get a large fraction of honest votes of sufficient depth. They will all vote for $H$ because $H$ remains the only block at its level, and extends the tip of a longest notarized chain. So there exists a $\dr$ such that   $\underbar{V}_B[s]$ exceeds $\frac{m}{2} +1$ for all $s \ge r+\dr$ and $B$ is notarized at round $r+\dr$ and beyond. By Lemma \ref{ass:half_votes}, the actual number of votes $V_B[s]$ will also be greater than $\frac{m}{2}+1$ for all $s > r+\dr$.

\epf

The following lemma is also needed in the liveness proof. 

\begin{lemma}
\label{lem:balance}
Assume  $\beta< 1/2$, $\bar{f}_v$ is chosen sufficiently small, and $m$ sufficiently large. Under the typical event $\texttt{T}$,  there exists a $\drtwo$, depending only on $\beta$, such that if a honest proposer block $H$ is broadcast at round $r$ and 
there is no other proposer block at a different level extending the tip of a longest notarized chain until round $r+\drtwo$, 
then no block on level $\ell(H)$ made public after round $r+\drtwo$ will be notarized in any future rounds after $r+\drtwo$.   
\end{lemma}

\bpf
Let $H$ be a honest proposer block that appears at round $r$. It is on the tip of the longest notarized chain. It follows from Lemma E.2 (macroscopic chain-growth), Lemma E.4 (macroscopic chain-quality) in \cite{prism} and $\beta < 1/2$ that a large fraction of voter chains grow and have positive chain quality. Hence, if one waits for a sufficient, but finite, many number of rounds, one can get a large fraction of honest votes of sufficient depth. They will all vote for $H$, or another proposer block at the same level of $H$, while no other proposer block at a different level and extending the tip of a longest notarized chain is made public after round $r$.  So there exists a $\drtwo$ such that after $r+\drtwo$,  there will always be more than $m/2$ voter chains each of which votes for a proposer block at level $\ell(H)$ that has been made public before round $r+\drtwo$, provided that no other proposer block at a different level and extending the tip of a longest notarized chain is made public after round $r$ and before round $r+\drtwo$. Thus, any proposer block that is made public at level $\ell(H)$ after round $r+\drtwo$ can at no round get more than $m/2$ votes. By Lemma  \ref{ass:half_votes}, this means that no proposer block made public at level $\ell(H)$ after round $r+\drtwo$ can be notarized. 
\epf

We note that both Lemma~\ref{lem:notarize} and Lemma~\ref{lem:balance} hold true when we increase $\dr$ and $\drtwo$ to $\max(\dr,\drtwo)$. WLOG, we assume that $\dr = \drtwo$ is used in Lemma~\ref{lem:notarize} and Lemma~\ref{lem:balance} henceforth.

We define the critical property about notarized blocks that will be needed in the security proof of \ps. 
Suppose we have two proposer blocks $B$ and $B'$. They have certified levels $\ell(B),\ell(B')$ on the proposer blocktree, and depths $d(B), d(B')$ in the notarized blokchain respectively.

Consider the ``bad'' event, defined for these two proposer blocks: Event $\E_{B,B'}$ holds if there exist rounds  $r$ and $r'$, such that $B$ is notarized in round $r$ and $B'$ is notarized in round $r'$. 
One property of interest is, defined for $B$ and $B'$: 

\begin{lemma}
\label{prop:weak}
Under the typical event $\texttt{T}$, the following is true over the entire horizon:  Given any two proposer blocks $B$ and $B'$, if (i) $\ell(B)  = \ell(B')$,  or (ii) $\ell(B) < \ell(B')$ and $d(B) > d(B')$, then the event $\E_{B,B'}$ cannot hold. 
\end{lemma}

\bpf
Let us assume the typical event $\texttt{T}$ holds. If the event $\E_{B,B'}$ holds, then by Lemma \ref{ass:half_votes}, there exists some round $r$ such that more than $\frac{m}{2}$ voter chains vote for $B$ and more than $\frac{m}{2}$ voter chains vote for $B'$. Hence there must be at least one voter chain which votes for both $B$ and $B'$. Since a voter chain can only vote for at most one proposer block on each level, this implies that $\ell(B) \not = \ell(B')$. Without loss of generality, let us assume that $\ell(B) < \ell(B')$. Then the voter block $v'$ that votes for $B'$ is a descendant of the block $v$ that votes for $B$. By the validity rule for voter chain, $d(B') \ge d(B)$. This proves the lemma. 
\epf





\begin{lemma}\label{property:level_order}
Under the typical event $\texttt{T}$, the following holds over the entire horizon: After a proposer block at level $\ell$ is notarized, no private proposer block at level $\ell' \leq \ell$ will ever be notarized.
\end{lemma}
\bpf
Assume the typical  event $\texttt{T}$ holds. Say a proposer block $B$ at level $\ell$ is notarized at round $r$, which means there is a set of voter blocks $\V_B(r)$ at round $r$ each of which is on the longest chain of a distinct voter chain and voted for $B$. By Lemma~\ref{ass:half_votes}, there is a subset of $\V_B(r)$, denoted by $\V'_B \subseteq \V_B(r)$ with $|\V'_B| > \frac{m}{2}$, will stay on the longest chain at any future round after $r$. We denote the set of indices of the voter chains for the blocks in $\V'_B$ as $\C'_B$. For each voter chain in $\C'_B$, a block voted for $B$ is on the longest chain at round $r$ and stays on the longest chain afterwards. Then, for a block $B'$ at level $\ell(B') \leq \ell(B)$ held in private at round $r$, first it must have not received any votes on the chains in $\C'_B$ since it is in private, second it will not receive any votes on the chains in $\C'_B$ any time in any future round since otherwise the block voting for $B'$ will be a child of the block voting for $B$, and the fact $\ell(B') \leq \ell(B)$ is clearly a violation of the voting rule. Therefore, the maximum number of votes $B'$ can have at any time after round $r$ is less than $\frac{m}{2}$, and it will never be notarized.
\epf

\begin{lemma}\label{property:order}
Under the typical event $\texttt{T}$, the following holds over the entire horizon. Say at round $r$ the tip of the public notarized tree has depth $d$, then any proposer block that appears after round $r$ with a depth $d' < d$ will not be notarized. \end{lemma}
\bpf
Assume the typical event $\texttt{T}$ holds. Suppose the deepest notarized block $B$ in round $r$ has depth $d$. This implies that at round $r$, there is a set of voter blocks $\V_B(r)$, each of which is on the longest chain of a distinct voter chain and voted for $B$. By Lemma~\ref{ass:half_votes}, a subset of $\V_B(r)$, denoted by $\V'_B \subseteq \V_B(r)$ with $|\V'_B| > \frac{m}{2}$, will stay on the longest chain at any future round after $r$. We denote the set of indices of the voter chains for the blocks in $\V'_B$ as $\C'_B$. On each chain in $\C'_B$, any voter block $v$ arriving after round $r$ will be a child of a block $v' \in \V'_B$ on the longest chain. The depth of the proposer block $v$ votes for cannot be less than $d$ since it can be inferred from $v'$ that the tip of the notarized proposer chain has already reached depth $d-1$ at an earlier round. Therefore, voter blocks that vote for depth $d' < d$ will never appear on the voter chains in $\C'_B$ after round $r$, a proposer block at depth $d'<d$ that appears after round $r$ can never receive more than $\frac{m}{2}$ votes and will never be notarized.
\epf



    


\subsection{Consistency of \ps}
Here we present our consistency theorem for the proposed \ps protocol.

\begin{theorem}[Consistency]
\label{thm:consistent}
For sufficiently large $m$, under the typical event $\texttt{T}$ with probability at least $1-\epsilon_m$, the following holds over the entire horizon. Let us assume an honest node observes a notarized proposer blockchain of depth $d+1$, with the last three trailing proposer blocks $B_{d-1}(\ell-1)$, $B_d(\ell)$, and $B_{d+1}(\ell+1)$ on consecutive blocktree levels $\ell-1$, $\ell$, and $\ell+1$. Then, it is impossible that some other honest nodes confirm a proposer block $B'_{d}(\ell')$ distinct from $B_d(\ell)$ at depth $d$ of a notarized chain, 
\end{theorem}

\bpf
The proof follows by directly applying Lemma~\ref{prop:weak} for different values of $\ell'$ under the typical event $\texttt{T}$. For $B'_{d}(\ell')$ to be confirmed at some honest node, it has to be notarized by that node. When $\ell' \leq \ell-1$, we apply Lemma~\ref{prop:weak} to $B_{d-1}(\ell-1)$ and $B'_{d}(\ell')$ to conclude that they cannot be both notarized. When $\ell' = \ell$, we know from Lemma~\ref{prop:weak} $B_{d}(\ell)$ and $B'_{d}(\ell')$ cannot be both notarized. Finally, when $\ell' \geq \ell+1$, we can see that $B_{d+1}(\ell+1)$ and $B'_{d}(\ell')$ cannot be both notarized. Hence, in all cases $B'_{d}(\ell')$ cannot be notarized, and hence cannot be confirmed.  \epf

\begin{remark}
Interestingly, the protocol is consistent for any $\beta$ between $0$ and $1$. However, if $\beta > 1/2$, the protocol never notarizes any blocks, because it knows that all the votes can be reverted by the powerful adversary, i.e. $\underbar{V}_B[r] = 0$ for all $r$ and for all $B$. Hence in this range, the protocol is not live.
\end{remark}

The question now is for which values of $\beta$ the protocol is live. We address this next. 

\subsection{Liveness of \ps}
We prove that \ps is live as long as $\beta < \frac{1}{2}$. That is, for all possible attack strategies, as long as the fraction of the adversary mining power $\beta < \frac{1}{2}$, any honest transaction submitted to the system will eventually be included in the confirmed ledger, after constant expected latency. Hence, \ps retains the optimal resilience of \bitcoin. 

\subsubsection{Liveness attack examples}
\label{app:liveness_attacks}
To appreciate the challenge of proving the liveness of \ps up to $\beta = 1/2$, we give here two examples of successful liveness attacks under {\em specific} arrival ordering of proposer blocks where the adversary has only $1/3$ of all the blocks. Even with so few blocks, the adversary can completely displace honest blocks from the longest notarized chain for \ps.

\begin{figure}[htbp]
    \centering
 \includegraphics[width=1\textwidth]{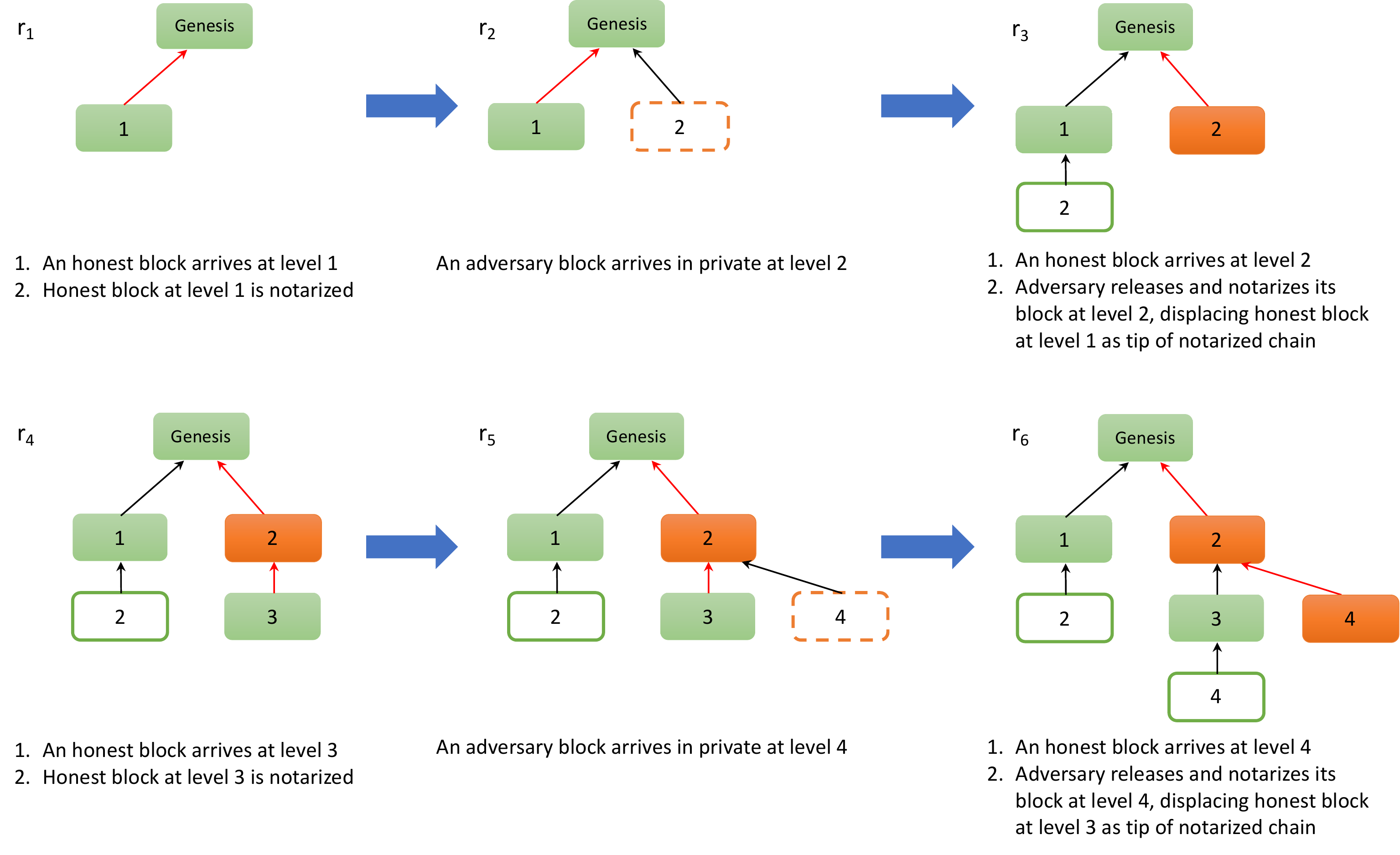}
 \caption{Illustration of the liveness attack for arrival sequence HAH HAH $\ldots$. Honest and adversary blocks are colored green and orange respectively. Private blocks have dashed borders, and the borders of public blocks are solid. Blocks become filled when they are notarized and left hallow otherwise. The height of each block indicates its depth on the notarized chain, and the links point to the depth parents. Links along the longest notarized chain in the current tree are colored red. Blocks are labelled by their levels on the proposer tree. An adversary block arrives in private in round $r_2$, and is released later in round $r_3$ to interrupt the notarization of an honest block arriving on the same level, and simultaneously notarized to displace another honest block out of the longest notarized chain. This can continue to occur as the HAH arrival pattern repeats, and no honest block will make into the longest notarized chain albeit having twice the number of adversary blocks.}
    \label{fig:live_attack1}
\end{figure}

\noindent {\bf Example 1: arrival sequence HAH HAH $\ldots$.} (Fig.~\ref{fig:live_attack1}) Proposer blocks arrive periodically in tuples of honest-adversary-honest blocks.  In this attack, each adversary proposer block simultaneous stops the notarization of an honest proposer block at the same level of the proposer tree and displaces another notarized honest proposer block at the same depth on the notarized chain. This ``one adversary block displaces two honest blocks'' action can continue to occur as the HAH arrival pattern repeats.

\begin{figure}[htbp]
    \centering
 \includegraphics[width=1\textwidth]{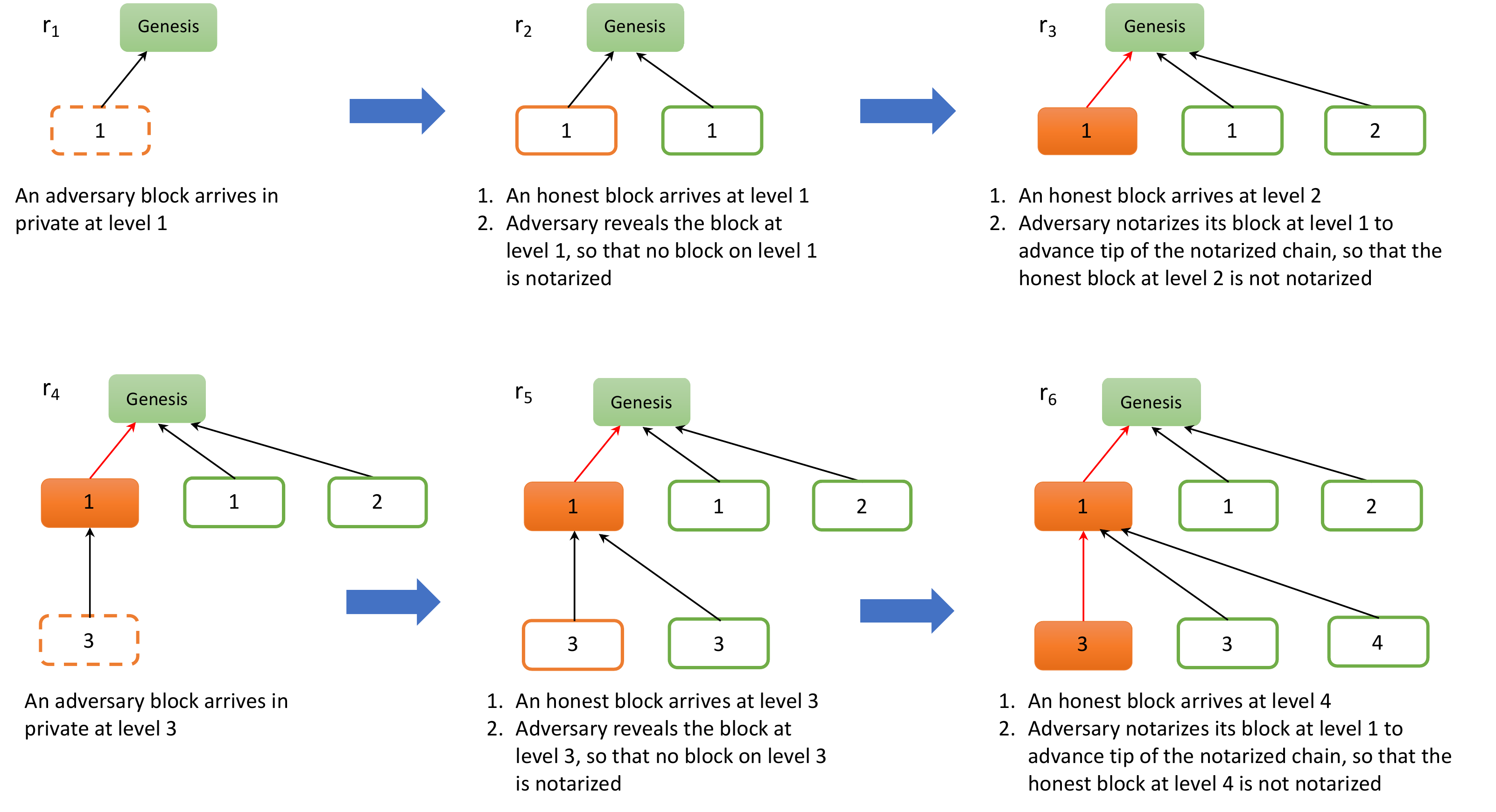}
 \caption{Illustration of the liveness attack for arrival sequence AHH AHH $\ldots$. Honest and adversary blocks are colored green and orange respectively. Private blocks have dashed borders, and the borders of public blocks are solid. Blocks become filled when they are notarized and left hallow otherwise. The height of each block indicates its depth on the notarized chain, and the links point to the depth parents. Links along the longest notarized chain in the current tree are colored red. Blocks are labelled by their levels on the proposer tree. Each adversary block arrives in private, and released to balance the votes on an honest block that arrives later on the same level (rounds $r_2$ and $r_5$). While none of these two blocks is immediately notarized, the adversary subsequently notarizes its (public) block to interrupt the notarization of anther honest block that later arrives at the same depth (in rounds $r_3$ and $r_6$). This action can continue to occur if the AHH arrival pattern repeats, making none of the honest block notarized.}
    \label{fig:live_attack2}
\end{figure}

\noindent {\bf Example 2: arrival sequence AHH AHH $\ldots$.} ( Fig.~\ref{fig:live_attack2}) Proposer blocks arrive periodically in tuples of adversary-honest-honest blocks. Similarly to the previous arrival sequence, with only half of the number of honest blocks, the adversary can launch an attack to make none of the honest blocks notarized. More specifically, each adversary block is released to interrupt the notarization of an honest block on the same level, and later notarized to interrupt the notarization of another honest block on the same depth. This ``one adversary block displaces two honest blocks'' action can continue to occur as the HAH arrival pattern repeats. 
 
\subsubsection{Liveness analysis}

One observation about the two attack examples is that the arrival pattern of the proposer blocks is {\em periodic}. This means that these patterns occur with vanishing probability over a long horizon. Thus, although the adversary only has $1/3$ of the blocks in these scenarios, these examples do not contradict the claim of liveness of \ps up to $\beta = 1/2$. This is because  liveness is a probabilistic notion, with the confirmation latency defined in (\ref{def:latency}) as the {\em average} latency. However, the examples do suggest that existing techniques to prove liveness for longest chain protocols will run into difficulties. In particular, the standard approach to prove liveness for the \bitcoin longest chain protocol (via the notions of chain quality and chain growth \cite{backbone}) is based on the technique of {\em block matching}: to displace one honest block from the longest chain, one adversary block is needed to match the honest block at the same level of the blockchain. Hence, to stop liveness, the adversary needs as many blocks as the honest miners have. When $\beta < 1/2$, the adversary will with high probability not have that many blocks, and hence the \bitcoin protocol is live. This argument cannot be directly used to prove liveness for \ps for all $\beta < 1/2$ because, as seen in the above attack examples, one adversary block can displace {\em two} honest blocks, matching one by level and one by depth. A direct block matching argument can only prove liveness up to at most $\beta = 1/3$. Can one prove liveness all the way up to $\beta = 1/2$?

 Looking deeper into the periodic patterns of the two attack scenarios, it can be seen that they share an important characteristic: after each period (in round $r_3$ and round $r_6$), the adversary doesn't have any more private blocks in store to attack in the future; it has to rely on more adversary blocks to come to continue the attack. The probabilistic nature of the block arrivals says that this periodic arrival pattern cannot be repeated forever; eventually there will be a burst of several consecutive honest proposer blocks, and their notarization cannot be stopped by the adversary.  We will show that this phenomenon holds in general: no matter what the attack strategy is, as long as $\beta < 1/2$, the system will with probability $1$ return infinitely often to a ``Genesis state" where the adversary has no more private blocks that can be utilized in the future to interrupt the notarization and confirmation of honest blocks. After returning for large enough number of times, there will eventually be a time in which there is a burst of consecutive honest blocks which will be notarized and confirmed. 

We formally define these concepts below.

\begin{definition}[Notarizability]
A proposer block  $B$ is said to be notarizable at round $r$ if it is possible that in some future round $r' \ge r$, $B$ will be notarized.
\end{definition}

\begin{example}
In Fig. \ref{fig:live_attack1}, the adversary block in round $r_2$ is notarizable, but the same block is no longer notarizable in subsequent rounds after it is notarized. In Fig. \ref{fig:live_attack2}, the adversary block in round $r_1$ is notarizable. The same block remains notarizable in round $r_2$ although it has been made public to balance the votes of the honest block. By reversing a very small number of voting chains in the future, the adversary can get the block notarized. 
\end{example}

\begin{definition}[BlockDAG]
The blockDAG $\DAG[r]$ at round $r$ is the directed 
acyclic graph whose nodes correspond to all the proposer blocks and all the voter blocks that have been mined until that round, honest or adversary, private or public. Directed edges between these nodes include parent links between voter blocks, votes from voter blocks to propose blocks, and parent links between proposer blocks (two types of parents: level and depth).
\end{definition}

\begin{definition}[Genesis state]
We say that the blockDAG $\DAG[r]$ at round $r$ is in the Genesis state, if there is no private adversary block which is notarizable and no private adversary proposer block at a greater level than all honest proposer blocks.
\end{definition}

\begin{example}
In Fig. \ref{fig:live_attack1}, the blockDAG is in the Genesis state in rounds $r_1,r_3,r_4,r_6$, and not in the Genesis state in rounds $r_2$ and $r_5$. In Fig. \ref{fig:live_attack2}, the blockDAG is in the Genesis state in rounds $r_2, r_3, r_5, r_6$ and not in the Genesis state in rounds $r_1$ and $r_4$.
\end{example}

The following is the key result which shows that the return to genesis state is inevitable for any attack strategy as long as $\beta < 1/2$.

\begin{theorem}
\label{thm:counting}
Assume $\beta< 1/2$, $\bar{f}_p$ and $\bar{f}_v$ are chosen sufficiently small, and $m$ sufficiently large. Under the typical event $\texttt{T}$ (with probability of at least $1-\epsilon_m$) and any adversary attack strategy, during the execution of the \ps protocol, assuming the blockDAG leaves the Genesis state in round $R_0 \geq 1$, and stay away from the Genesis state up to round $R$, or more explicitly $\DAG[R_0-1]$ is in the Genesis state, and none of $\DAG[R_0], \DAG[R_0+1],\ldots, \DAG[R]$ is in the Genesis state, it must hold that 
\begin{equation}
\label{eq:matching}
M_a(R_0,R) \geq M_h(R_0,R)-1,
\end{equation}
where $M_a(R_0,R)$ and $M_h(R_0,R)$ are respectively adversary and honest blocks that arrive between rounds $R_0$ and $R$.
\end{theorem}


\begin{example}
In Fig. \ref{fig:live_attack1}, in round $r_2$, a notarizable adversary block arrives in private, moving $\DAG[r_2]$ out of Genesis state. In this case, we have $M_a(r_2,r_2)=1$ and $M_h(r_2,r_2)=0$, which is consistent with Theorem \ref{thm:counting}. 
\end{example}

Theorem \ref{thm:counting} implies that the return to Genesis state is inevitable if $\beta < 1/2$, since in that case the mining rate of adversary blocks is less than the mining rate of honest blocks, and with high probability the condition (\ref{eq:matching}) cannot hold for large $R- R_0$. 

The two attack examples imply that it is not true  the number of adversary blocks have to be larger than the number of honest blocks over the {\em entire} execution of the protocol if it is not live. Theorem \ref{thm:counting}, however, says that it {\em is } true during the periods when the blockDAG is not in the Genesis state. This enables us to still use a block matching argument, but confined to these periods. However, the block matching argument to prove Theorem \ref{thm:counting} is significantly more involved than the argument to prove the liveness of \bitcoin. Despite the fact that an adversary block can simultaneously denotarize two honest blocks, one at the same level and one at the same depth, we show that if that happens when the blockDAG is not in a genesis state, then there must be another adversary block which remains notarizable.    

The proof of Theorem \ref{thm:counting} is given in the next subsection. In the meantime, let us turn to what happens when we get back to Genesis state, so that we can complete the proof of liveness.

\begin{lemma}\label{lem:chain_growth}
Assume $\beta< 1/2$, $\bar{f}_v$ is chosen sufficiently small, and $m$ sufficiently large. Under the typical event $\texttt{T}$, when the blockDAG is in the Genesis state, and the next three arriving proposer blocks $B_1$, $B_2$, and $B_3$ are all honest with inter-arrival times of more than $\dr$ rounds, then denoting their levels $\ell_i$ and depths $d_i$ for $i=1,2,3$ respectively, we have
\begin{itemize}
    \item $\ell_2 = \ell_1+1$, and $\ell_3 = \ell_2+1$.
    \item $d_2 = d_1 + 1$, $d_3 = d_2+1$, and $B_2$ and $B_3$ will be both notarized.
    \item No conflicting block $B' \neq B_2$ at depth $d_2$ will ever be notarized.
\end{itemize}
\end{lemma}
\bpf
We denote the level of the honest block who has the greatest level among all honest blocks when $B_1$ arrives as $L_h$. Since the blockDAG is in the Genesis state, no private block will have a level greater than $L_h$, and hence all private blocks have levels strictly smaller than $\ell_1$ when $B_1$ arrives. $B_1$, $B_2$, and $B_3$ will be on consecutive levels according to the block mining rule. 

We denote the depth of the tip of the notarized chain when $B_1$ arrives as $D$. We observe that in round $r(B_1)$, there are no public blocks with depth $d > D+1$  since no block on depth $D+1$ is notarized yet. Also since the block tree is in the Genesis state, no private block can be made public to interrupt the notarization of $B_1$. When $B_1$ arrives at depth $d_1 = D+1$, within $\dr$ rounds, it is either notarized, or not notarized due to adversary notarizing some block made public before round $r(B_1)$ at depth $d_1$. For either case, the tip of the notarized chain advances to depth $d_1$. When $B_2$ arrives at $d_2 = d_1+1$, there is no public block and notarizable private block on $d_2$, and $B_2$ will be notarized. Similarly, $B_3$ will be notarized at depth $d_3 = d_2+1$. Finally, after $B_3$ is notarized, according to Lemma~\ref{property:order}, no blocks made public in the future on depth $d_2 < d_3$ will ever be notarized, and $B_2$ becomes the only notarized block on depth $d_2$ in the notarized chain.
\epf

\begin{lemma}
\label{lem:four_block}
Assume $\beta< 1/2$, $\bar{f}_v$ is chosen sufficiently small, and $m$ sufficiently large. Under the typical event $\texttt{T}$, when the blockDAG is in the Genesis state, and the next four arriving blocks $B_1$, $B_2$, $B_3$, and $B_4$ are honest with inter-arrival times of more than $\dr$ rounds, then block $B_3$, together with its notarized prefix chain, will be confirmed.  
\end{lemma}
\bpf
Since the blockDAG is in the Genesis state when $B_1$ arrives, and $B_1$ is honest, the blockDAG will stay in the Genesis state when $B_2$ arrives. Applying Lemma~\ref{lem:chain_growth} to the block tuples $(B_1, B_2, B_3)$ and $(B_2, B_3, B_4)$ respectively, we have that all these four blocks will be on consecutive levels, as well as consecutive depth. Blocks $B_2$, $B_3$, and $B_4$ will be notarized with $B_2$ and $B_3$ being the only notarized block on their respective depths. Hence $B_2 \leftarrow B_3 \leftarrow B_4$ forms a notarized chain of three blocks with consecutive levels, we can confirm $B_3$ together with its notarized prefix chain according to the confirmation rule.
\epf

This leads to the liveness theorem for \ps.

\begin{theorem}[Liveness]
\label{thm:live}
Let $\beta < 1/2$. Then there exist sufficiently small $\bar{f}_v$,$\bar{f}_p$ such that \ps is live, with finite worst case expected $\epsilon_m$-latency, where $\epsilon_m$ (defined in (\ref{eqn:constants})) decreases exponentially in the number of voter chains $m$. 
\end{theorem}
\bpf
Assume the typical event $\texttt{T}$ holds and  $\bar{f}_v$, $\bar{f}_p$ chosen sufficiently small. Suppose a transaction ${\sf tx}$ arrives at round $r_0$. Let $T_0$ be the number of rounds it takes to first return to a Genesis state starting at round $r_0$ ($T_0=0$ if $\DAG[r_0]$ is already at Genesis state). After round $r_0 + T_0$, we define two random variables $X_1$ and $T_1$ such that 1) if the next proposer block that arrives is an honest block, $X_1 = 1$ and $r_0+T_0+T_1$ is the arrival round of that block; 2) if the next proposer block that arrives is a public adversary block, or a private adversary block which is neither notarizable nor on a level greater than all honest blocks in the blockDAG, $X_1 = 2$ and $r_0+T_0+T_1$ is the arrival round of that block; 3) if the next proposer block that arrives is an adversary private block which is either notarizable or has a greater level than all honest blocks in the blockDAG, $X_1 = 3$ and $r_0+T_0+T_1$ is the first round that the blockDAG returns to the Genesis state. Note that in all three cases, the blockDAG is in Genesis state at round $r+T_0+T_1$
Similarly, we define $T_2, T_3, \ldots$ and random variables $X_2,X_3 \ldots$ such that for all $i$, the blockDAG is in Genesis state at round $r_0+T_0+T_1 +\ldots + T_i$. By Lemma \ref{lem:four_block}, the confirmation latency of ${\sf tx}$ is bounded by
\begin{equation}
\label{eq:stopping}
    \sum_{i=0}^N T_i
\end{equation}
where $N$ is the stopping time defined as the smallest $n$ such that $X_n = X_{n-1} = X_{n-2} = X_{n-3} = 1$, i.e. the first time we get $4$ honest block arrivals in a row after reaching Genesis state. This is because by this time, a new honest proposer block is confirmed, and the transaction ${\sf tx}$ will appear either in this block or an earlier confirmed block.

Now, for $i \ge 1$, let us define $Y_i = 1$ if $X_i = 1$ and $Y_i = 0$ otherwise. Then
\begin{eqnarray*}
    \E[T_i|Y_i = 1] & = & \frac{1}{(1-\beta)\bar{f}_p}\\
    \E[T_i|Y_i = 0] & = & \frac{1}{\beta \bar{f}_p} + \E[T_i'|X_i = 3]\textup{Pr}(X_i = 3|Y_i = 0) \\
    & \le & \frac{1}{\beta \bar{f}_p} + \E[T_i'|X_i=3]
\end{eqnarray*}
where $T_i'$ is the number of rounds to return to Genesis state after a private notarizable adversary block arrives. Using Theorem \ref{thm:counting}, $\E[T_i'|X_i = 3]$ is bounded by the expected time a random walk $\{S[r]\}$ first crosses $-1$ starting from $S[0] = +1$, where the random walk increments by $1$ every time there is an adversary arrival and decrements by $1$ every time there is a honest arrival. Since $\beta < 1/2$, the drift of this random walk is negative, and hence the expected hitting time is finite.  Hence, one can find a constant $c_1(\beta)$, depending only on $\beta$, such that for all $i > 1$:
\begin{equation}
\label{eq:exp_bound}
    \E[T_i|Y_i] < c_1(\beta).
\end{equation}
Substituting into (\ref{eq:stopping}), we get the following bound on the average latency:
\begin{eqnarray*}
    \E[\sum_{i=0}^N T_i] & = & \E[\E[\sum_{i=0}^N T_i|N]]\\
    & = & \E[T_0] + \E[\sum_{i=1}^N E[T_i|N]|N]\\
    & = & \E[T_0] + \E[\sum_{i=1}^N E[T_i|Y_i]|N] \label{eq:markov}\\
    & \le& \E[T_0] + \E[N]c_1(\beta),
\end{eqnarray*}
where (\ref{eq:markov}) comes from the fact that $T_i, Y_i,N$ forms a Markov chain, and the last inequality comes from (\ref{eq:exp_bound}).

$\E[N]$ is equal to the expected number of independent coin flips until getting $4$ consecutive Heads, where the probability of getting a Head is $1-\beta$. This is finite, depending only on $\beta$. Thus, it remains only to show that $\E[T_0]$ is also finite.

Let $X_0 = 0$ if the blockDAG is in Genesis state at round $r_0$ (when the transaction arrives). Let $X_0 =1$ if otherwise. If $X_0 = 0$, then $T_0=0$. Hence, it remains only to bound $\E[T_0|X_0=1]$. Let $R_0$ be the earliest round such that the blockDAG is not in Genesis state for all rounds $R_0,R_0+1, \ldots r_0$. Let us define a random walk, $\{W[r]\}$, starting at $r = r_0$, such that 
$$ W[r_0] \triangleq M_a(R_0,r_0) - M_h(R_0,r_0),$$
and $W[r]$ increments by $1$ whenever an adversary block arrives, and decrements by $1$ whenever a honest block  arrives. By Theorem \ref{thm:counting}, $\E[T_0|X_0 = 1]$ is upper bounded by the expected time for the random walk $\{W[r]\}$ to first hit $-1$. Now let us look at the distribution of $W[r_0]$ at the initial round $r_0$:
\begin{eqnarray*}
    & & \textup{Pr}(W[r_0] \ge  w) \\
    & = & \textup{Pr}( M_a(R_0,r_0) - M_h(R_0,r_0) \ge w)\\
    & \le & \textup{Pr}( \max_{0\le r \le r_0} M_a(r,r_0) - M_h(r,r_0) \ge w)
\end{eqnarray*}
The random process in the last probability can be interpreted as another random walk $\{w'[r]: r=r_0, r_0-1,\ldots,0\}$, starting at round $r_0$ at the origin (i.e., $w'[r_0]=0$) and running in reverse and crosses the level $w$ at some (earlier) round. This random walk also increments by $+1$ whenever an adversary block arrives and decrements by $-1$ whenever a honest block arrives, and hence it also has negative drift. Hence, the level crossing probability decreases exponentially with $w$. Thus, the  distribution of $W[r_0]$ has an exponential tail. Moreover, the expected first hitting time conditioned on $W[r_0]=w$ is linear in $w$ for large $w$. Hence, the expected hitting time is finite. This proves that $\E[T_0]$ is finite, thus completing the liveness proof. 
\epf

Our proof of liveness of \ps has some parallelism with the proof of liveness of \streamlet \cite{streamlet}. There it is proved that a new block is finalized whenever there are $8$ honest leaders in consecutive epochs. Finalization needs $6$ consecutive honest epochs. The first few additional honest epochs are needed to ``undo the damage that corrupt leaders have done", in the language of the authors. This is roughly analogous to the notion of ``return to Genesis state" in our proof of liveness of \ps. However, unlike \streamlet, a finite number of honest proposer levels is not sufficient to guarantee to ``undo adversarial damage" in \ps. This is because in \ps , the adversary can publish proposer blocks that were mined privately and arbitrarily earlier to attempt to disrupt liveness. This cannot happen in \streamlet because honest nodes only vote for proposer blocks proposed in the current epoch. This difference is a consequence of the use of the levels of a proposer chain instead of absolute time to mark epochs in \ps, to enable dynamic availability.

Despite the differences, return to Genesis state can still be proved for \ps, albeit with a significantly complex proof. We next turn to this.

\subsubsection{A block-matching proof of Theorem \ref{thm:counting}}

This subsection is devoted to proving the key result Theorem \ref{thm:counting}, which shows that the return to the Genesis state is inevitable as long as $\beta < 1/2$. First we need some basic definitions and facts. 

\begin{definition}[Loner blocks]
We say that an honest proposer block $H$ that arrives at round $r$ is a loner if there is no other honest proposer blocks arrived or arriving between rounds $r-\dr$ and $r+\dr$, where $\dr$, as defined in Lemma~\ref{lem:notarize}, is the maximum number of rounds before which we expect $H$ to be notarized.
\end{definition}


We choose the mining rate of proposer blocks $\bar{f}_p \ll 1$ sufficiently small such that the number of proposer blocks that arrive in a round can be approximated as a Bernoulli random variable with success probability $\bar{f}_p$, and the inter-arrival times of proposer blocks are i.i.d. geometric with success rate of $\bar{f}_p$. The probability that two consecutively arriving proposer blocks are more than $\dr$ rounds apart is $P(\dr) = (1-\bar{f}_p)^{\dr} \approx e^{-\bar{f}_p \dr}$ for small $\bar{f}_p$. Here we choose the ratio between the mining rate of the voter blocks on a single voter chain and the mining rate of the proposer blocks $\frac{\bar{f}_v}{\bar{f}_p} \gg 1$, such that the expected inter-arrival time of proposer blocks $\frac{1}{\bar{f}_p}$ is much larger than $\dr = \Theta(\frac{1}{\bar{f}_v})$ rounds. This choice leads to $\bar{f}_p \dr \ll 1$, and we can make $P(\dr)$ arbitrarily close to $1$. From now on we focus on the scenario where over the entire horizon, all blocks arrive at least $\dr$ rounds apart from each other, and consequently all honest blocks are loners. 

\begin{figure}[htbp]
    \centering
 \includegraphics[width=\textwidth]{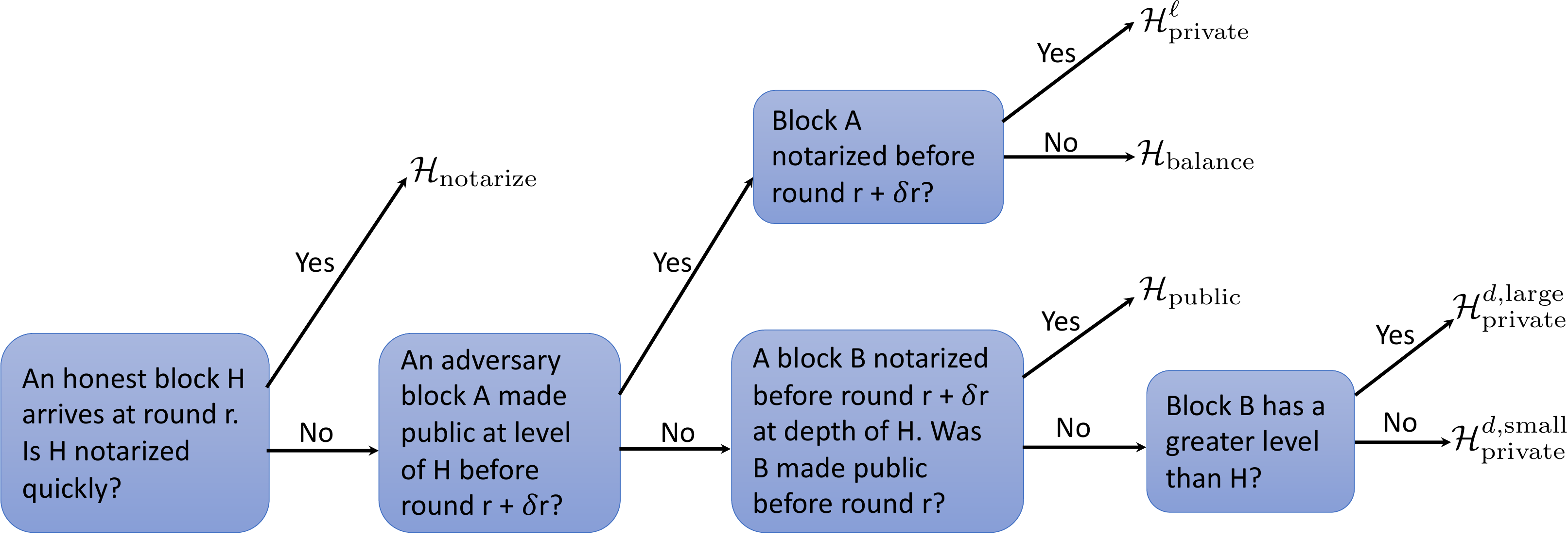}
 \caption{A decision tree to partition the set of honest blocks $\H$. Note that by Lemma~\ref{lem:notarize}, when an honest block is not quickly notarized and no adversary block made public at the same level within $\dr$ rounds, the tip of the notarized chain must have been advanced to match the depth of the honest block.}
    \label{fig:partition_half}
\end{figure}

When an honest (loner) block $H$ arrives at round $r$, and gets notarized before round $r+\dr$, we say $H$ is notarized \emph{quickly}. We know from Lemma~\ref{lem:notarize} that $H$ will be notarized quickly unless there is another block made public at the level of $H$, or notarized at the depth of $H$ to advance the length of the longest notarized chain from its length at round $r$. 

\begin{definition}\label{def:partition}
Assume $\beta< 1/2$, $\bar{f}_p$ and $\bar{f}_v$ are chosen sufficiently small, and $m$ sufficiently large. Under the typical event $\texttt{T}$, we define a partition of the set of honest blocks $\H$ in a blockDAG. Specifically, as illustrated in Fig.~\ref{fig:partition_half}, for an honest block $H$ in the blockDAG arrived in round $r(H)$ on level $\ell(H)$ and depth $d(H)$, it will fall into one of the following mutually exclusive subsets. 
\begin{align*}
\H_{\textup{notarize}} &\triangleq \{H \textup{ is quickly notarized}\}, \\
\H^{\ell}_{\textup{private}} &\triangleq \{\textup{adversary publishes and notarizes a block at level $\ell(H)$ before} \\
&\textup{round $r(H)+\dr$}\},\\
\H_{\textup{balance}} &\triangleq \{\textup{adversary publishes but not notarizes a block at level $\ell(H)$ before}\\ 
&\textup{round $r(H)+\dr$}\},\\
\H_{\textup{public}} &\triangleq \{\textup{adversary notarizes a block made public before round $r(H)-\dr$}\\ &\textup{to extend longest notarized chain by one from its length at round $r(H)$,}\\
&\textup{before round $r(H)+\dr$}\},\\
\H^{d,\textup{small}}_{\textup{private}} &\triangleq \{\textup{adversary publishes and notarizes a block $B$ with level $\ell(B) < \ell(H)$}\\ &\textup{to extend longest notarized chain by one from its length at round $r(H)$,}\\
&\textup{before round $r(H)+\dr$}\},\\
\H^{d,\textup{large}}_{\textup{private}} &\triangleq \{\textup{adversary publishes and notarizes a block $B$ with level $\ell(B) > \ell(H)$}\\ &\textup{to extend longest notarized chain by one from its length at round $r(H)$,}\\
&\textup{before round $r(H)+\dr$}\}.
\end{align*}
We also further partition $\H_{\textup{public}}$ into $\H_{\textup{public}} = \H^h_{\textup{public}} \cup \H^a_{\textup{public}}$. For each $H \in \H_{\textup{public}}$, we denote the block made public before round $r(H)-\dr$ and notarized between rounds $r(H)$ and $r(H)+\dr$ at the depth of $H$ as $B$ ($d(B) = d(H)$). We classify an honest block $H$ in $\H_{\textup{public}}$ into $\H^h_{\textup{public}}$, if $B \in \H_{\textup{balance}}$, or $B$ is the adversary block made public between rounds $r(H')$ and $r(H')+\dr$ to stop the notarization of some honest block $H' \in \H_{\textup{balance}}$ arrived before $H$. Otherwise, $H$ is classified into $\H^a_{\textup{public}}$. The block $B$ for $H \in \Hpa$ has to be adversary.
\end{definition}

\begin{example}
In Fig. \ref{fig:live_attack1}, the honest block that arrives in round $r_1$ is notarized quickly and belongs to $\Hn$; the block that arrives in round $r_3$ sees an adversary block notarized on the same level (level 2) before round $r_3+\dr$ and hence belongs to $\Hpl$. In Fig.~\ref{fig:live_attack2}, the honest block that arrives in round $r_2$ sees an adversary block made public on the same level (level 1) before round $r_2+\dr$ and no block is notarized on that level after round $r_2+\dr$, hence the honest block belongs to $\Hb$; in round $r_3$, an honest block arrives on level 2 and sees an adversary block, made public before round $r_3$, notarized at the same depth (depth 1) before round $r_3+\dr$. Since this adversary block was made public on level 1 to balance the votes of a previous honest block arrived in round $r_2 < r_3$ on level 1, the honest block that arrives in round $r_3$ belongs to $\Hph$.
\end{example}

Assume $\beta< 1/2$, $\bar{f}_p$ and $\bar{f}_v$ are chosen sufficiently small, and $m$ sufficiently large. With probability of at least $1-\epsilon_m$ ($\epsilon_m$ defined in (\ref{eqn:constants})), the following facts hold for the blockDAG generated running \ps under the typical event $\texttt{T}$.

\begin{fact}\label{fact:no_grow}
For each $H \in \H^h_{\textup{public}}$, we denote the block notarized at the same depth as $H$ between rounds $r(H)$ and $r(H)+\dr$ as $B$. The block $B$ can be either $H'$ or $A'$, where $H'$ is some honest block in $\Hb$ arrived before $H$ whose notarization was interrupted by some adversary block $A'$ made pubic at the same level $\ell(A') = \ell(H')$ between rounds $r(H')$ and $r(H') + \dr$. Then $H$ and $H'$ must be on the same depth of the notarized chain. That is, the longest notarized chain cannot grow between rounds $r(H')$ and $r(H)$.
\end{fact}
\bpf
Since $H'$ arrives before $H$, we have $d(H) \geq d(H')$. Now let us assume $d(H) > d(H')$. In this case, $B = A'$ since $d(B) = d(H)$. Now we have $d(H) = d(A') > d(H')$. Recall that $A'$ was made public on the same level as $H'$ to balance $H'$'s votes within $\dr$ rounds of $H'$'s arrival. For $A'$ to be considered a valid block upon publication with $d(A') > d(H')$, it must have a notarized parent $P$ at depth $d(H')$, which did not exist before $H'$ arrived (since honest nodes mine on the tip of the notarized chain). Therefore, either $P$ was made public and notarized between rounds $r(H')$ and $r(H')+\dr$, or $P$ was public before round $r(H')$ and notarized after $H'$ arrived. Consequently, $H'$ should have been classified into $\Hpl$, $\Hpds$, $\Hpdl$, or $\H_{\textup{public}}$, but not $\H_{\textup{balance}}$. This yields a contradiction, and hence we must have $d(H) = d(H')$.
\epf

\begin{example}
In Fig.~\ref{fig:live_attack2}, we denote the honest block that arrives in round $r_3$ as $H$. The block $H$ belongs to $Hph$ and its corresponding $H'$ is the honest block that arrived in round $r_2$ on level 1 with $H' \in \Hb$. We verify that $H$ and $H'$ are on the same depth, i.e., $d(H) = d(H') =1$, which is consistent with Fact~\ref{fact:no_grow}.
\end{example}

\begin{fact}\label{fact:self_balance}
For each $H \in \H^a_{\textup{public}}$, we denote the adversary block, made public before round $r(H) - \dr$ and notarized at the same depth as $H$ between rounds $r(H)$ and $r(H)+\dr$ as $B$. When $B$ was made public in round $r$, there must be at least one other adversary block $A \neq B$ that was made public on level $\ell(A) = \ell(B)$ between rounds $r-\dr$ and $r+\dr$. Moreover, the block $A$ has to be notarizable when it was made public. 
\end{fact}
\bpf
Suppose otherwise $B$ was the only notarizable adversary block on level $\ell(B)$ between rounds $r-\dr$ and $r+\dr$. There are two possible situations. 1) $B$ was the only notarizable block on level $\ell(B)$ between rounds $r-\dr$ and $r+\dr$. Say the greatest level on which a block is notarized in $\DAG[r]$ is $L[r]$, and the tip of the notarized chain in $\DAG[r]$ has depth $D[r]$. As $B$ is later notarized at a depth $d(B) = d(H) > D[r]$, we know by Lemma~\ref{property:level_order} that the level $\ell(B) > L[r]$. Therefore, $B$ will be notarized quickly when it was made public, before $H$ arrives. 2) There was another honest block $H'$ on level $\ell(B)$ between rounds $r-\dr$ and $r+\dr$, and $H'$ arrives before round $r$. First, by definition of $H^a_{\textup{public}}$, $H' \notin \H_{\textup{balance}}$. Also, $H'$ cannot be from $\H_{\textup{notarize}}$, $\H^{\ell}_{\textup{private}}$, or $\Hpdl$ since otherwise $B$ would have not been notarizable after round $r+\dr$. Finally, if $H'$ were from $\H^{d,\textup{small}}_{\textup{private}}$ or $\H_{\textup{public}}$, $H'$ will not receive any vote on level $\ell(H') = \ell(B)$, and $B$ would have been notarized quickly.
\epf

\begin{fact}\label{fact:notarize}
For each honest block $H \in \H_{\textup{notarize}}$, which is notarized in some round $r$ between $r(H)$ and $r(H)+\dr$, if $\DAG[r]$ is not in the Genesis state, there must be some private adversary block $A$ arrived before round $r(H)$ at level $\ell(H)$.
\end{fact} 
\bpf
Assuming otherwise, $H$ would be the only block at level $\ell(H)$ when it arrives, and all private adversary blocks would be on levels strictly smaller than $\ell(H)$. By Lemma~\ref{property:level_order}, the notarization of $H$ in round $r$ would render none of these private blocks notarizable, and the blockDAG will enter the Genesis state.
\epf

We also note that since the private adversary block $A$ is on the same level of $H$, the notarization of $H$ will make $A$ never notarizable, even when made public in a later round.

\begin{fact}\label{fact:exist}
For each $H \in \H^h_{\textup{public}} \cup \H^{d,\textup{small}}_{\textup{private}}$, we denote the block notarized at the same depth as $H$ in some round $r$ between $r(H)$ and $r(H)+\dr$ as $B$. It must be the case $\ell(B) < \ell(H)$. If $\DAG[r]$ is not in the Genesis state, an adversary block $A$ that falls into one of the following two categories must exist:
\begin{itemize}
    \item Category 1: $A$ arrived before round $r(H)$ in private on level $\ell(H)$,
    \item Category 2: $A$ arrived before round $r(H)$ in private on level $\ell(B) < \ell(A) < \ell(H)$, and remains in private and notarizable in round $r$.
\end{itemize}
\end{fact}
\bpf
If $H \in \H^h_{\textup{public}}$, since $B$ was made public before $r(H)$, we have $\ell(H) > \ell(B)$ according to the mining rule. If $H \in \Hpds$, we have by definition of $\Hpds$ that $\ell(H) > \ell(B)$.
If there is no block arrived in private on level $\ell(H)$ before round $r(H)$, $H$ would be the only block on level $\ell(H)$ in round $r(H)$ and all private blocks will be on levels strictly smaller than $\ell(H)$. Now since $\DAG[r]$ is not in Genesis state, there exists at least one private notarizable block $A$ whose level $\ell(A)$, by Lemma~\ref{property:level_order}, has to be greater than $\ell(B)$.
\epf

\begin{definition}\label{def:hidden}
For an honest block $H \in \H^h_{\textup{public}} \cup \H^{d,\textup{small}}_{\textup{private}}$, and the block $B$ that is notarized at the same depth of $H$ in round $r$ between $r(H)$ and $r(H)+\dr$, if $\DAG[r]$ is not in the Genesis state, we denote the set of adversary blocks in Category 1 in Fact~\ref{fact:exist} as $\A^1_H$, and the set of adversary blocks in Category 2 in Fact~\ref{fact:exist} as $\A^2_H$. We know by Fact~\ref{fact:exist} that $\A^1_H$ and $\A^2_H$ cannot be both empty. Particularly, we define the following adversary block $A(H)$ for $H$.
\begin{align*}
    A(H) = \begin{cases}
    \textup{earliest arrived block in $\A^1_{H}$}, & \A^1_{H} \neq \emptyset,\\
    \textup{earliest arrived block on the level $\min\{\ell(A): A \in \A^2_{H}\}$}, & \textup{otherwise}.
    \end{cases}
\end{align*}
\end{definition}

\noindent {\bf Proof of Theorem \ref{thm:counting}}

\noindent
\bpf
As the blockDAG exits the Genesis state in round $R_0$, it must be the case that some adversary block arrives in round $R_0$ in private and is notarizable in $\DAG[R_0]$. Since the next block arrives in at least $\dr$ rounds later (due to designed mining rates and the resulting loner assumption), the statement of the theorem holds trivially for $R < R_0+\dr$. In what follows, we focus on the case where $R \geq R_0 + \dr$.

We prove this theorem by constructing an injective map $\varphi$ from the set of honest blocks that arrive between rounds $R_0$ and $R$, denoted by $\H$, to the set of adversary blocks in $\DAG[R]$, denoted by $\A$, and demonstrating that the adversary blocks in the image $\varphi(\H)$ also arrive between rounds $R_0$ and $R$ (except for at most one of them).

We partition $\H$ as shown in Definition~\ref{def:partition} into seven subsets $\Hn$, $\Hpl$, $\Hb$, $\Hpdl$, $\Hpds$, $\Hph$, and $\Hpa$. We first define the map $\varphi$ for the subsets $\Hn$, $\Hpl$, $\Hb$, $\Hpdl$ as follows.
\begin{itemize}
    \item $H \in \H_{\textup{notarize}}$, $\varphi(H) =$ earliest arrived private adversary block before round $r(H)$ on level $\ell(H)$. Such block must exist according to Fact~\ref{fact:notarize};
    \item $H \in \H^{\ell}_{\textup{private}}$, $\varphi(H) =$ adversary block made public and notarized between rounds $r(H)$ and $r(H)+\dr$ at level $\ell(H)$;
    \item $H \in \H_{\textup{balance}}$, $\varphi(H) =$ adversary block $A$ made public between rounds $r(H)$ and $r(H)+\dr$ at level $\ell(H)$ and neither of $H$ and $A$ is notarized before round $r(H)+\dr$;
    \item $H \in \Hpdl$, $\varphi(H) =$ adversary block made public and notarized between rounds $r(H)$ and $r(H)+\dr$ at depth $d(H)$ and a level greater than $\ell(H)$.
\end{itemize}
We denote $\H_L \triangleq \Hn \cup \Hpl \cup \Hb \cup \Hpdl$, and argue that the adversary blocks in the image of $\H_L$ under the map $\varphi$, denoted by $\varphi(\H_L)$, each has a distinct level. First, according to the mining rule of \ps, for any two honest blocks $H$ and $H'$ in $\H$ with $r(H) < r(H')$, we have $\ell(H) < \ell(H')$ and $d(H) \leq d(H')$ in $\DAG[R]$. For two distinct blocks $H$ and $H'$ in $\H_L$ with $r(H) < r(H')$, we have either $\varphi(H)$ was made public before round $r(H)+\dr$ or $\ell(\varphi(H)) = \ell(H)$ for $H \in \Hn$. In either case we have $\ell(\varphi(H)) < \ell(H') \leq \ell(\varphi(H'))$.

Next we define the map $\varphi$ for the subsets $\Hpds$ and $\Hph$. Suppose there are $k$ honest blocks in $\Hpds \cup \Hph$, and we denote them as $\Hpds \cup \Hph = \{H_1,\ldots, H_k\}$ where $r(H_1) < \cdots < r(H_k)$ and hence $\ell(H_1) < \cdots < \ell(H_k)$. For each $H_i$, a block, denoted by $B_i$, is notarized at the same depth as $H_i$ between rounds $r(H_i)$ and $r(H_i) + \dr$ to extend the longest notarized chain in the blockDAG. Hence we have $d(H_1) < \cdots < d(H_k)$. We denote the adversary block associated with $H_i$ as defined in Definition~\ref{def:hidden} as $A_i = A(H_i)$. We know by Fact~\ref{fact:exist} that $\ell(B_i) < \ell(A_i) \leq \ell(H_i)$. We present some properties of $B_i$ and $A_i$ in the following.
\begin{description}
    \item \emph{Property 1:} for any $H_i \in \Hph$, $B_i$ can be either honest such that $B_i \in \Hb$, or adversary such that $B_i \in \varphi(\Hb)$. 
    \item \emph{Property 2:} for any $H_i \in \Hpds$, the level of $B_i$ is different from the level of any block in $\varphi(\H_L)$. To see this, consider two honest blocks $H_i \in \Hpds$ and $H_i' \in \H_L$. If $r(H_i) < r(H_i')$, we have $\ell(B_i) < \ell(H_i) < \ell(H_i') \leq \ell(\varphi(H_i'))$; if $r(H_i) > r(H_i')$, we know by Lemma~\ref{lem:notarize} and Lemma~\ref{lem:balance} that no more private blocks can be notarized on level $\ell(\varphi(H_i'))$ by the round of $r(H_i)$, hence we must have $\ell(B_i) > \ell(\varphi(H_i'))$.
    \item \emph{Property 3:} for any $H_i \in \Hpds \cup \Hph$, the level of $A_i$ is different from the level of any block in $\varphi(\H_L)$. For two honest blocks $H_i \in \Hpds \cup \Hph$ and $H_i' \in \H_L$, if $r(H_i) < r(H_i')$, we have $\ell(A_i) \leq \ell(H_i) < \ell(H_i') \leq \ell(\varphi(H_i'))$; if $r(H_i) > r(H_i')$, we know from the above property that $\ell(A_i) > \ell(B_i) > \ell(\varphi(H_i'))$.
\end{description}
For each $H_i$ in $H_1,\ldots,H_k$, we set $\varphi(H_i) = B_i$ or $A_i$. We do this successively as follows.

\emph{Step 1: initialization.} For $i=1$, if $H_1 \in \Hpds$, we set $\varphi(H_1) = B_1$; if $H_1 \in \Hph$, we set $\varphi(H_1) = A_1$.

\emph{Step 2: iterative assignment.} Suppose we have assigned $\varphi$ for $H_1,\ldots,H_m$, $1 \leq m \leq k$, and the following conditions hold.

\begin{description}
\item \emph{Condition 1:} $\varphi(H_1), \ldots, \varphi(H_m)$ are all on distinct levels;

\item \emph{Condition 2:} the levels $\ell(\varphi(H_1)), \ldots, \ell(\varphi(H_m))$ are different from the levels of the adversary blocks in $\varphi(\H_L)$;

\item \emph{Condition 3:} there exists at most one adversary block $A_m^p$ such that
    \begin{enumerate}
    \item $A_m^p = \varphi(H_j)$ for some $1 \leq j \leq m$;
    \item $\ell(B_m) < \ell(A_m^p)$.
    \end{enumerate}
We set $A_m^p = \emptyset$ if such block does not exist.
\item \emph{Condition 4:} if such $A_m^p$ exists, its level $\ell(A_m^p) \leq \ell(H_m)$.
\end{description}

By Properties 2 and 3 above we can verify that these conditions hold for $m=1$. Specifically, we have $A_1^p = \emptyset$ if $H_1 \in \Hpds$, and $A_1^p = A_1$ if $H_1 \in \Hph$. For the next honest block $H_{m+1}$, we describe our assignment of $\varphi(H_{m+1})$ in different cases.

\begin{figure}[htbp]
    \centering
 \includegraphics[width=\textwidth]{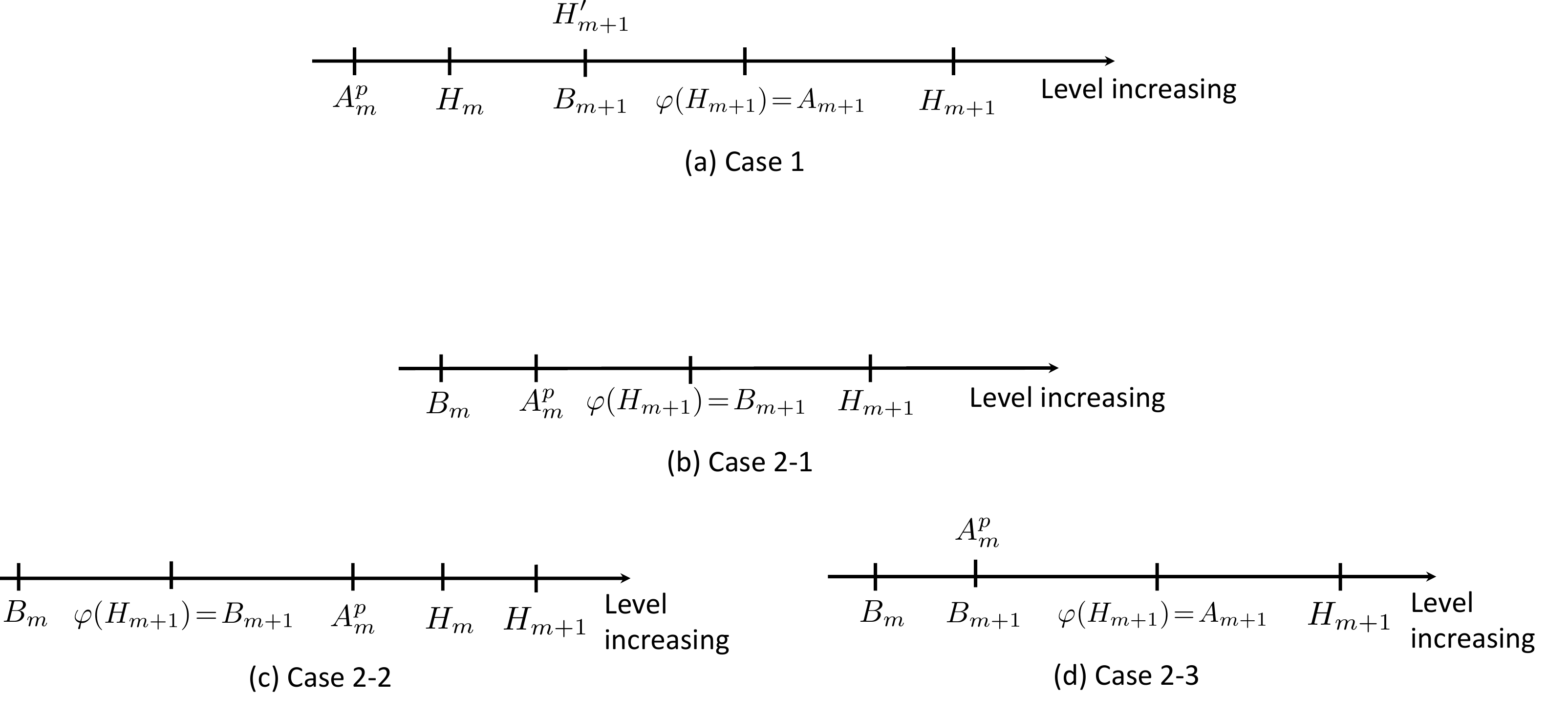}
 \caption{Illustration of the orderings of the levels of relevant blocks in different cases for $H_{m+1}$.}
    \label{fig:cases}
\end{figure}

\emph{Case 1: $H_{m+1} \in \Hph$.} In this case, $B_i$ can be either honest or adversary (see Property 1). We define $H_{m+1}' = B_i$ if $B_i$ is honest, or $H_{m+1}' = \varphi^{-1}(B_i)$ if $B_i$ is adversary. $H_{m+1}'$ is in $\Hb$ and $\ell(H_{m+1}') = \ell(B_{m+1})$. We know from Fact~\ref{fact:no_grow} that $d(H_{m+1}') = d(H_{m+1})$. Also, since $d(H_{m+1}) > d(H_m)$, we have $d(H_{m+1}') > d(H_m)$, and hence $H_{m+1}'$ arrives later than $H_m$ and $\ell(H_{m+1}') > \ell(H_m)$. The ordering of the levels of these blocks is illustrated in Fig.~\ref{fig:cases}(a). In this case, we set $\varphi(H_{m+1}) = A_{m+1}$. The above four conditions continue to hold for $H_1,\ldots,H_{m+1}$. To see this, Condition 1 holds as $\ell(\varphi(H_{m+1})) > \ell(B_{m+1}) > \ell(H_m) \geq \max(\ell(\varphi(H_1)),\ldots,\ell(\varphi(H_m)))$; Condition 2 holds by the virtue of Property 3; Conditions 3 and 4 hold as there exists a unique $A_{m+1}^p = \varphi(H_{m+1})$.

\emph{Case 2: $H_{m+1} \in \Hpds$.} In this case, $B_{m+1}$ is made public and notarized between rounds $r(H_{m+1})$ and $r(H_{m+1})+\dr$ on level $\ell(B_{m+1}) < \ell(H_{m+1})$. Since $B_{m+1}$ is private and notarizable in round $r(H_{m+1}) > r(H_m)+\dr$ when $B_m$ was already notarized, we have by Lemma~\ref{property:level_order} that it has to be the case that $\ell(B_{m+1}) > \ell(B_m)$. We further consider the following three sub-cases.

\emph{Case 2-1: $\ell(B_{m+1}) > \ell(A_m^p)$ or $A_m^p = \emptyset$.} The ordering of the levels of relevant blocks is illustrated in Fig.~\ref{fig:cases}(b). In this sub-case, we set $\varphi(H_{m+1}) = B_{m+1}$. All four conditions continue to hold for $H_1,\ldots,H_{m+1}$. To see this, Condition 1 holds as $\ell(\varphi(H_{m+1})) > \ell(A_m^p) \geq \max(\ell(\varphi(H_1)),\ldots,\ell(\varphi(H_m)))$ if $A_m^p$ exists, or $\ell(\varphi(H_{m+1})) > \ell(B_m) \geq \max(\ell(\varphi(H_1)),\ldots,\ell(\varphi(H_m)))$ otherwise; Condition 2 holds due to Property 2. Conditions 3 and 4 hold as $A_{m+1}^p = \emptyset$.

\emph{Case 2-2: $\ell(B_{m+1}) < \ell(A_m^p)$.} The ordering of the levels of relevant blocks is illustrated in Fig.~\ref{fig:cases}(c). In this sub-case, we set $\varphi(H_{m+1}) = B_{m+1}$. All four conditions continue to hold for $H_1,\ldots,H_{m+1}$. To see this, Condition 1 holds as there is only one block in $\varphi(H_1), \ldots, \varphi(H_m)$, i.e., $A_m^p$ has level greater than $B_m$, and $\ell(B_m) < \ell(\varphi(H_{m+1})) < \ell(A_m^p)$; Condition 2 holds due to Property 2. Condition 3 holds as there exists a unique $A_{m+1}^p = A_m^p$, and condition 4 holds as $\ell(A_{m+1}^p) \leq \ell(H_m) < \ell(H_{m+1})$.

\emph{Case 2-3: $\ell(B_{m+1}) = \ell(A_m^p)$.} The ordering of the levels of relevant blocks is illustrated in Fig.~\ref{fig:cases}(d). In this sub-case, we set $\varphi(H_{m+1}) = A_{m+1}$. All four conditions continue to hold for $H_1,\ldots,H_{m+1}$. To see this, Condition 1 holds as $\ell(\varphi(H_{m+1})) > \ell(A_m^p) \geq \max(\ell(\varphi(H_1)),\ldots,\ell(\varphi(H_m)))$; Condition 2 holds due to Property 3. Conditions 3 and 4 hold as there exists a unique $A_{m+1}^p = \varphi(H_{m+1})$.

Now we have obtained an assignment of the map $\varphi$ on all blocks $H_1,\ldots,H_k$ in $\Hpds \cup \Hph$ such that the above four conditions hold for $m=k$. 
Finally, we define the map $\varphi$ on the honest blocks in $\Hpa$. For each $H \in \H^a_{\textup{public}}$, we denote the adversary block made public in round $r < r(H) - \dr$ and notarized at the same depth as $H$ between rounds $r(H)$ and $r(H)+\dr$ as $B$. We know from Fact~\ref{fact:self_balance} that there must exist another adversary block $A \neq B$ that was made public on level $\ell(A) = \ell(B)$ between rounds $r-\dr$ and $r+\dr$. Also, the block $A$ has to be notarizable when it was made public. Now, if the adversary block $B$ is not in $\varphi(\H_L \cup \Hpds \cup \Hph)$, we set $\varphi(H) = B$; otherwise, we set $\varphi(H) = A$, and we know $A \notin \varphi(\H_L \cup \Hpds \cup \Hph)$ as $\ell(A) = \ell(B)$ and distinct blocks in $\varphi(\H_L \cup \Hpds \cup \Hph)$ are on distinct levels in $\DAG[R]$. At this point, we have constructed a map $\varphi$ from the set $\H$ of honest blocks that arrive between rounds $R_0$ and $R$ to the set $\A$ of adversary blocks in $\DAG[R]$, and demonstrated that $\varphi$ is injective. Next we proceed to show that out of the set of adversary blocks in $\varphi(\H)$, at most one of them arrived before round $R_0$, given that $\DAG[R_0-1]$ is in the Genesis state.

For an honest block $H \in \H_L$, $\varphi(H)$ is private in round $r(H)$. If $\varphi(H)$ arrived before round $R_0$, it has to be private in round $R_0-1$. We denote the greatest level of all honest blocks in $\DAG[R_0-1]$ as $L_h[R_0-1]$, and  we have $L_h[R_0-1] < \ell(H) \leq \ell(\varphi(H))$. Now $\varphi(H)$ is a private block in round $R_0-1$ on a level that is greater than all honest blocks in $\DAG[R_0-1]$, and this contradicts with $\DAG[R_0-1]$ being in the Genesis state. Hence, $\varphi(H)$ has to arrive in or after round $R_0$. For $H_i \in \Hpds \cup \Hph =  \{H_1,\ldots,H_k\}$, if $\varphi(H_i) = B_i$, we know $H_i \in \Hpds$ and $B_i$ is private and notarizable in round $r(H_i)$. If $\varphi(H_i)$ arrived before $R_0$, it has to be private and notarizable in round $R_0-1$, which contradicts with $\DAG[R_0-1]$ being in the Genesis state. On the other hand, if $\varphi(H_i) =A_i$, we know by Definition~\ref{def:hidden} that $A_i$ is private in round $r(H_i)$, with either $\ell(A_i) = \ell(H_i)$, or $\ell(A_i) < \ell(H_i)$ and $A_i$ being notarizable in round $r(H_i)$. In either case, as we have shown before it will lead to a contradiction with $\DAG[R_0-1]$ being in the Genesis state. Therefore, $\varphi(H_i)$ has to arrive in or after round $R_0$ for all $i=1,\ldots,k$. 

Finally, for an honest block $H \in \Hpa$, we denote the adversary block $B$ notarized between rounds $r(H)$ and $r(H)+\dr$ on the same depth of $H$ as $B$, and another adversary block $A$ made public between rounds $r-\dr$ and $r+\dr$ on the same level of $B$, where $r < r(H)$ is the round when $B$ was made public. 

\emph{Case 1: $B$ arrives before round $R_0$.} $B$ has to be public in round $R_0-1$ as $\DAG[R_0-1]$ is in the Genesis state. Thus $d(B) \leq D[R_0-1]+1$ where $D[R_0-1]$ is the length of the longest notarized chain in $\DAG[R_0-1]$. Also, since $d(B) = d(H) \geq D[R_0-1]+1$, we have $d(B) = D[R_0-1]+1$. In this case, since $r(B) < R_0$, $B \notin \varphi(\H_L \cup \Hpds \cup \Hph)$ and we have $B = \varphi(H)$. We note that there is at most one such block $B$ as the tip of the notarized chain increments after $B$ is notarized, and there is no public block with depth larger than $D[R_0-1]+1$ in $\DAG[R_0-1]$. 

\emph{Case 2: $B$ arrives in or after round $R_0$.} We show that the block $A$ cannot arrive before $R_0$. Let us assume otherwise, then $A$ has to be made public in a round $r' \leq R_0-1$, and by Fact~\ref{fact:self_balance} we know $B$ is made public in round $r < r' + \dr \leq R_0+\dr-1$. Therefore, $B$ has to be the private adversary block that arrives in round $R_0$ (i.e., $r(B) = R_0$) as the next block will arrive in at least $\dr$ rounds later than $R_0$. Now we can see that in round $R_0+\dr-1$, $B$ is the only block in $\DAG[R_0+\dr-1]$ that arrives in or after round $R_0$, and as it was made public in round $r < R_0+\dr-1$, $\DAG[R_0+\dr-1]$ is in the Genesis state. This contradicts with the assumption that the blockDAG stays away from Genesis state between rounds $R_0$ and $R \geq R_0 + \dr$. Therefore, both $B$ and $A$ arrive in or after round $R_0$, and $\varphi(H)$ arrives in or after round $R_0$.

To conclude, out of all the adversary blocks in $\varphi(\H)$, at most one of them in $\varphi(\Hpa)$ arrives before round $R_0$. Since we have shown that $\varphi$ is in injective, we must have $M_a(R_0,R) + 1 \geq M_h(R_0,R)$, where $M_a(R_0,R)$ and $M_h(R_0,R)$ are respectively adversary and honest blocks that arrive between rounds $R_0$ and $R$.
\epf

\bibliographystyle{plain}
\bibliography{references}


\end{document}